\documentclass{aa}
\usepackage{graphicx}
\usepackage{txfonts}
\usepackage{natbib}
\bibpunct{(}{)}{;}{a}{}{,}
\begin{document}

\title{Ground-based detection of Calcium and possibly Scandium and Hydrogen in the atmosphere of HD 209458b}

\titlerunning{Ground-based detection of Ca I and possibly Sc II and H I in the atmosphere of HD 209458b}

\author{N. Astudillo-Defru \inst{1,2} \and P. Rojo \inst{1}}

\institute{Departamento de Astronom\'ia, Universidad de Chile, Camino El Observatorio 1515, Las Condes, Santiago, Chile\\ \email{nicola.astudillo@obs.ujf-grenoble.fr} \and UJF-Grenoble 1 /CNRS-INSU, Institut de Plan\'etologie et d'Astrophysique de Grenoble (IPAG) UMR 5274, Grenoble, F-38041, France\\
}

\abstract
{Since the first exoplanetary atmosphere detection using the Hubble Space Telescope, characterization of exoplanet atmospheres from the ground have been playing an increasingly important role in the analysis of such atmospheres thanks to the enhancement of telluric correction techniques. At present, several species have been discovered in the atmosphere of HD 209458b, all of them consistent with theoretical models.}
{Data acquired using the High Dispersion Spectrograph on the Subaru telescope are re-analysed. We expect to discover new species in the atmosphere of the exoplanet HD 209458b. In addition to shed light on the atmospheric composition, we will derive the radial extension of the absorbents present in the atmosphere of the exoplanet.}
{We present an alternative method to correct the telluric effects through the analysis of variations in spectral lines with the airmass. To search absorptions due to an exoplanetary atmosphere we implemented an algorithm to automatically search for all the features presenting an atmospheric signature in the transmission spectrum and through the wavelength range in the data. In order to estimate uncertainties we perform a bootstrapping analysis.}
{Absorption excess due to the transitions of Calcium at 6162.17 {\AA} and 6493.78 {\AA}, Scandium at 5526.79 {\AA}, Hydrogen at 6562.8 {\AA} and Sodium doublet are detected in the transmission spectrum at a level of $-0.079\pm0.012\%$, $-0.138\pm0.013\%$, $-0.059\pm0.012\%$, $-0.123\pm0.012\%$, $-0.071\pm0.016\%$ using pass-bands of 0.5 {\AA}, 0.4 {\AA}, 0.5 {\AA}, 1.1 {\AA} and 0.6 {\AA}, respectively.}
{Models predict strong absorption in the Sodium resonance doublet which was previously detected, also in this analysis. However, this is the first report of Calcium and possibly Scandium in HD 209458b, including the possible ground-based detection of Hydrogen. Calcium is expected to condense out in the atmosphere of this exoplanet; therefore, confirmation of these results will certainly imply a review of theoretical models.}


\maketitle

\begin{figure*}[t]
\centering
\makebox[\textwidth]{
\includegraphics[scale=0.29]{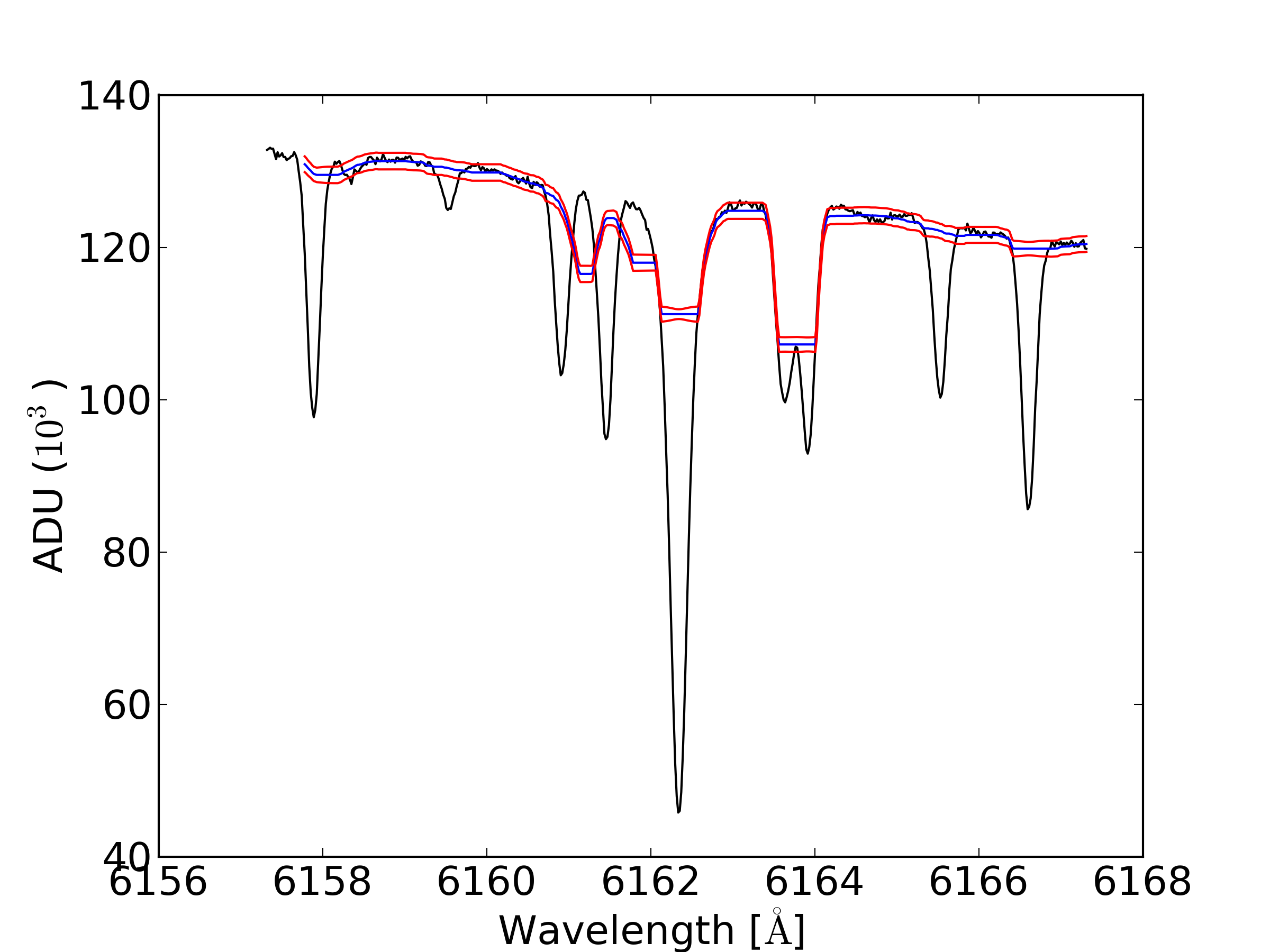}
\includegraphics[scale=0.29]{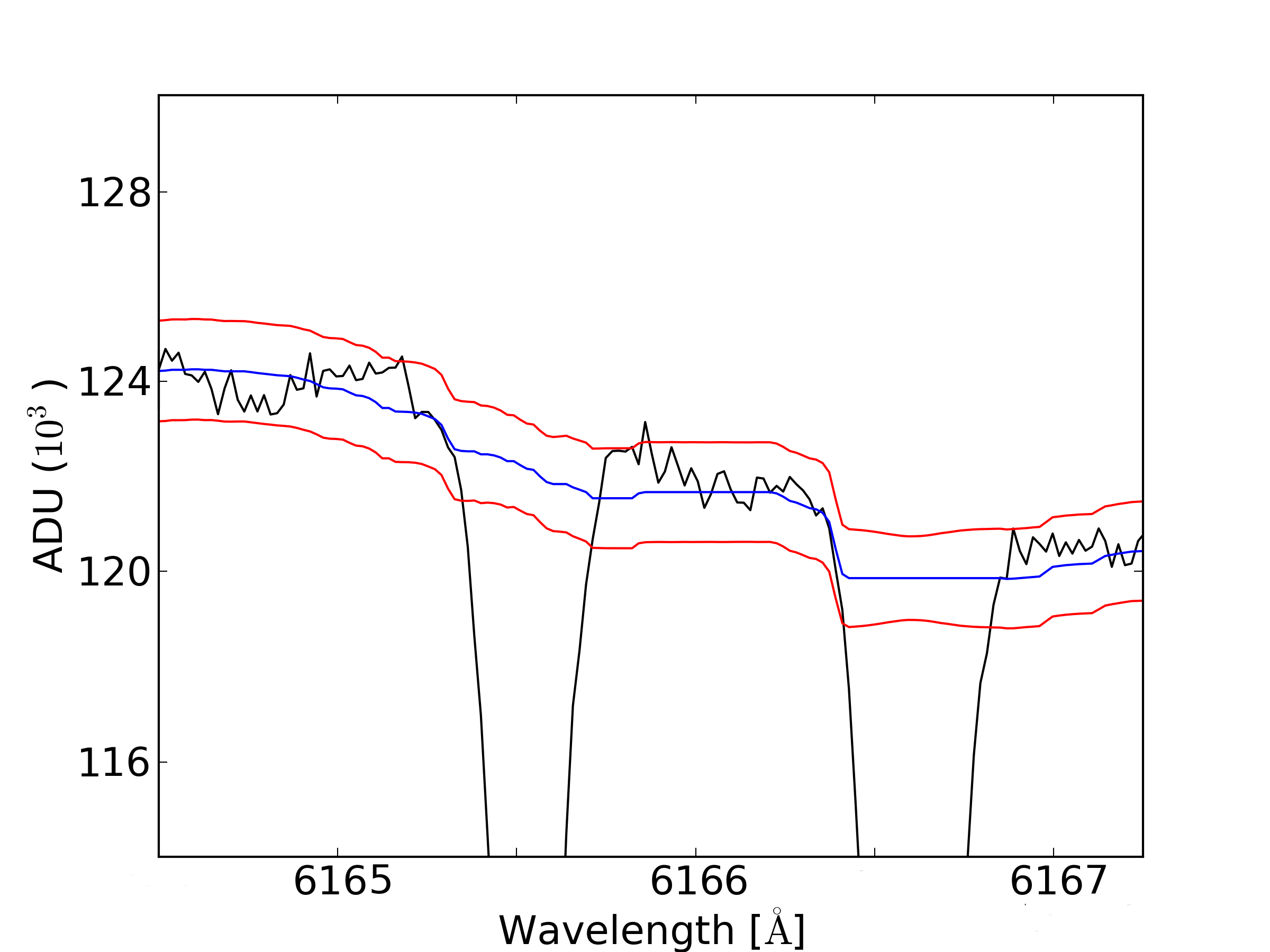}
\includegraphics[scale=0.29]{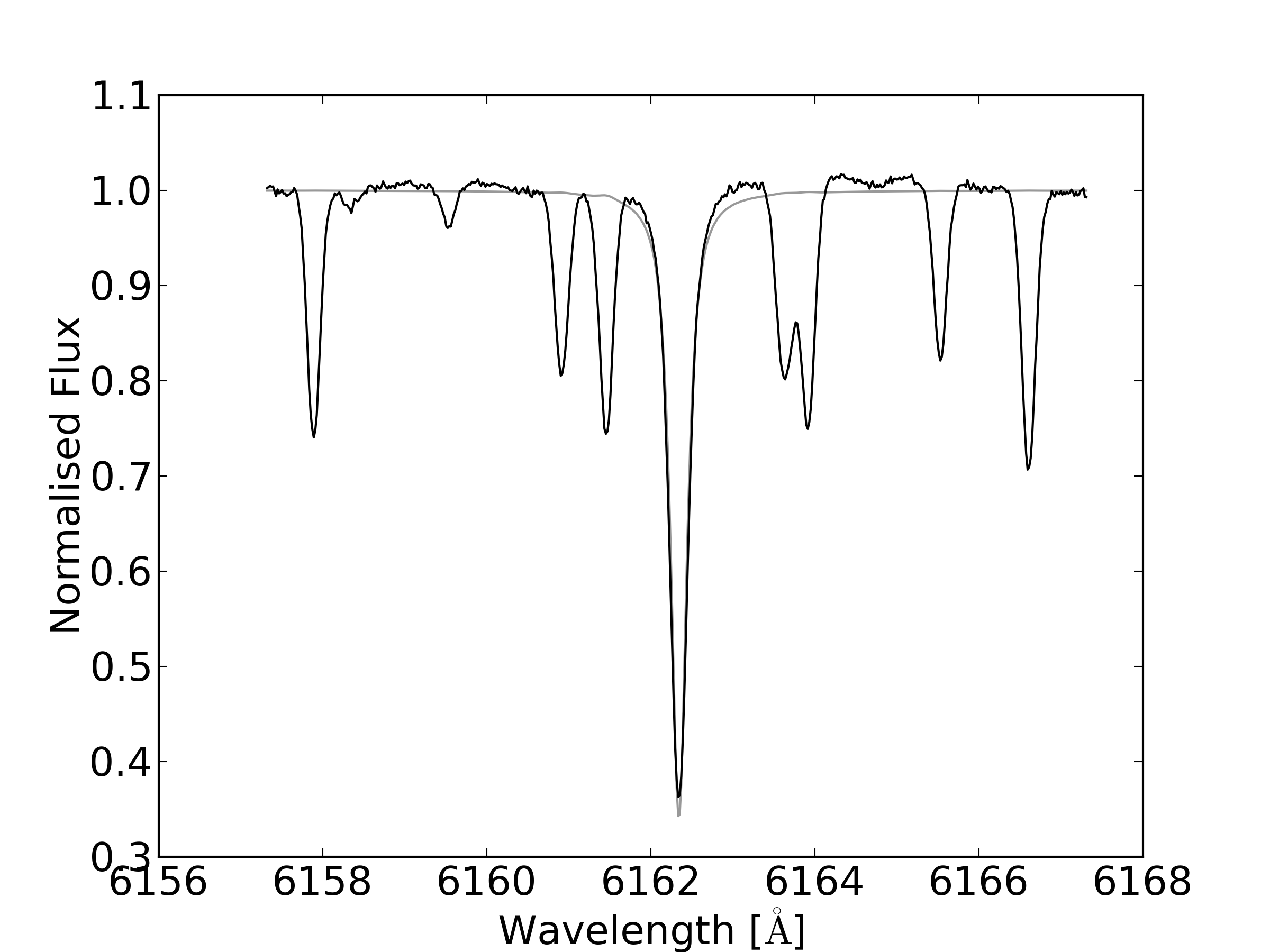}}
\caption{Local normalization around a spectral line. First (left plot), we obtain a median filtered array (middle blue line), the filter width was chosen to be 1.25 times the full-width-at-half-maximum (FWHM) - typically $FWHM \sim 0.4$ {\AA} - which corresponds to 50 pixels given the instrument resolution. Pixels within $3 \times FWHM$ and which intensity are beyond 3 Poisson-statistics sigmas from the median filter (red curves) are discarded. {\em Middle plot} shows a zoom to the median filter. The accepted pixels were used to fit a third order polynomial. \emph{Right:} The normalized spectrum after division by the polynomial and the Voigtian fit (gray) to the stellar line.}
\label{fig:norm}
\end{figure*}

\section{Introduction}

Since the first detection of an extrasolar planet \citep{Mayor_Queloz1995} radial velocities and transits monitoring have allowed the measurement of dynamics and bulk properties of hundreds of exoplanets. The former method is responsible of the majority of known planets and give us the minimum mass $m sin(i)$, among other parameters. The latter allowed 1/3 of known planets, adds thousand of candidates and provides the planet radius and inclination. Both methods are complementary, e.g. letting us to know the true planet mass. Moreover, transits present an unique opportunity to characterize exoplanets. Transmission spectroscopy allows to measure the absorption of radiation passing through the optically thin exoplanet atmosphere and decode the composition and physical parameters of the outer atmosphere (\citet{Charbonneau2002_first_atm_detection}, thereafter C2002).

Absorption from exoplanetary atmospheres are expected to be of the order of $~10^{-3} $ for a hot-Jupiter planet orbiting a Sun-like star. Therefore high signal-to-noise data are needed to measure such absorption. Regarding photon collecting capacity, ground-based telescopes takes advantage over space-based. However, data acquired using ground-based instruments must be corrected in order to deal with variations introduced by Earth's atmosphere. In previous work, these corrections are mainly done by observation of a reference star (simultaneously or not) or the creation of a synthetic telluric spectrum (\citet{Bean2010,Redfield2008_Na_ground_detection, Langland-Shula2009_keck+HIRES}, thereafter R2008 and L2009 respectively.)

Traditionally, theoretical models have been used to predict which transitions are prominent enough to be easily distinguishable from noise \citep{Burrows2010_atm_model, Fortney2010_TransSpec_model}. Detected components in the transiting HD 209458b's atmosphere agree with these models (C2002, \citet{Vidal-Madjar2003_lymanAlpha, Vidal-Madjar2004_O_C_HD209458b, Beaulieu2010_H2O_HD209458b, Desert2008_TiO_VO_HD209458b}). We note that previous detections were reported in a single spectral line (or doublet) or photometric detection instead of a suite of transitions of a particular specie. Although much work has been done to date, more studies need to be conducted to confirm some of the previous results and observation-models agreements, as well as improve techniques for precise constraining of abundances.

The purpose of present study was to perform a ``blind search'' in the whole transmission spectrum of HD 209458b to automatically detect strong and missed absorption produced in the exoplanetary atmosphere for every strong transition reported in databases. This research propose an alternative to the problem of correcting for non desired telluric effects on transmission spectroscopy. The topic is specially important in the search for biomarkers in exoplanetary atmospheres.

\section{Observations, data reduction and analysis}

Science archive SMOKA \citep{Baba2002Smoka} was used to obtain data collected with the High Dispersion Spectrograph (HDS) on the Subaru telescope. These HD209458 data have been described in \citet{Winn2004_HD209458_Halpha,Narita2005_HD209458_trans_spec,Snellen2008_Na_ground_detection}, thereafter W2004, N2005 and S2008 respectively. The entrance slit was $0.8''$ and the resolution was $R\sim45000$.

After applying the S2008 correction\footnote{$F_{cor}=F_{raw}-3.0 \times 10^{-6} \times F_{raw}^2$} for the CCD non-linearity effects we followed the HDS IRAF Reduction Manual available on the NAOJ website\footnote{http://www.naoj.org/Observing/Instruments/HDS/}. We used 29 spectra frames each with a total exposure time of 500s. The resulting signal to noise ratio (SNR) varies between 300 and 450 per pixel in the continuum. We concentrated in the red CCD that contains 21 orders covering from 5500{\AA} to 6800{\AA}.

\citet{Knutson2007_HD209458b_properties} ephemeris was used to select the in-transit and out-of-transit frames. We independently and automatically analyzed the region around each of the transitions indexed in The Interactive Database of Spectral Standard Star Atlases\footnote{http://spectra.freeshell.org/spectroweb.html} \citep{2008JPhCS.130a2015L}. This database lists the line parameters, including rest wavelength, oscillator strength ($g_if_{ij}$), expected line depth, and responsible element for telluric and stellar transitions for a subset of template standard stars. For HD 209458 (G0 V) we use the solar spectrum template because is the closest spectral type. Using this database limit our study to transitions present in the host star, i.e. we are restricted to exoplanet elements that are also present in the parent star atmosphere.

\subsection{Data preparation}
\label{sec:data_preparation}

\begin{figure*}[t]
\centering
\makebox[\textwidth]{
\includegraphics[scale=0.29]{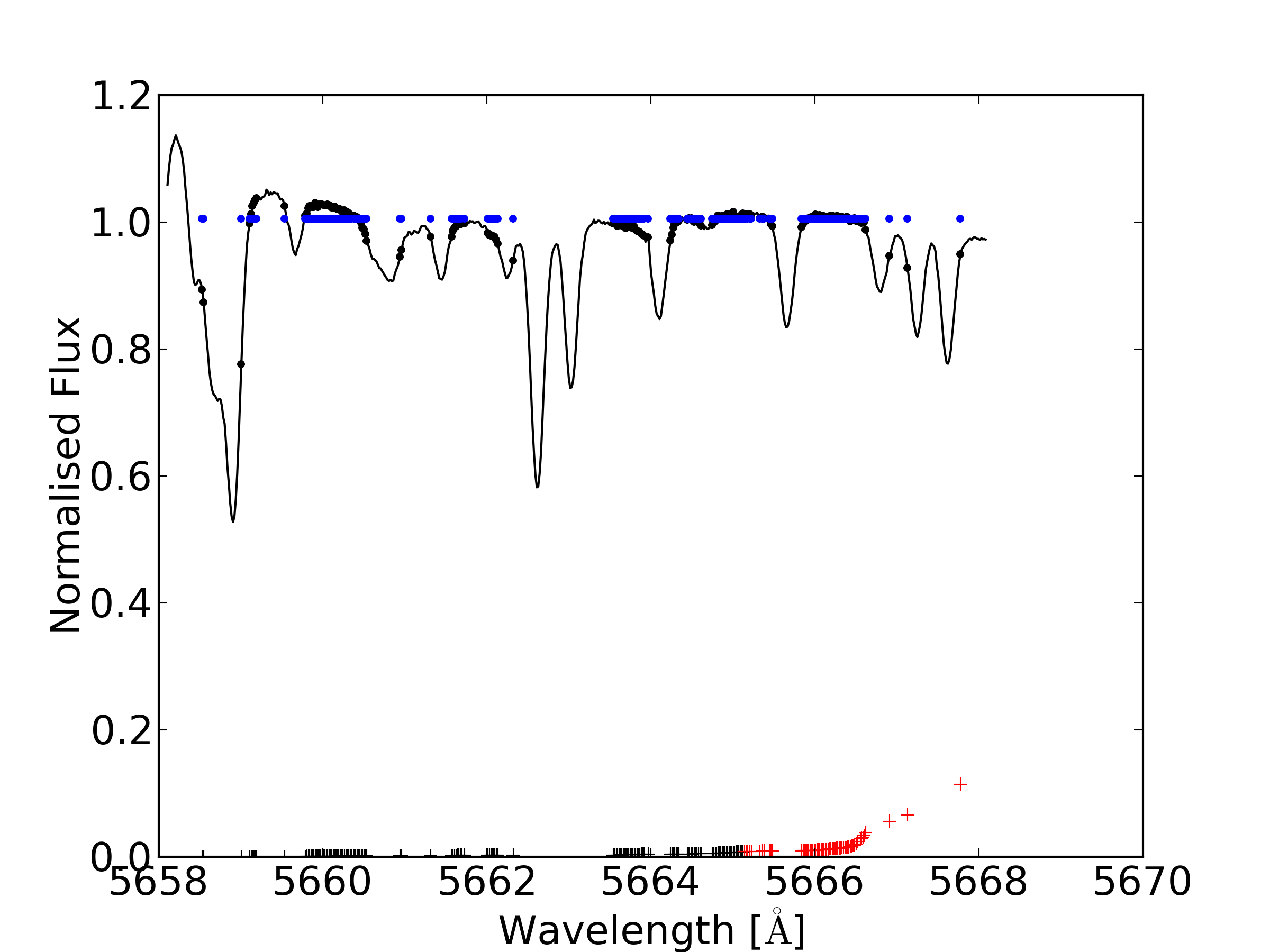}
\includegraphics[scale=0.29]{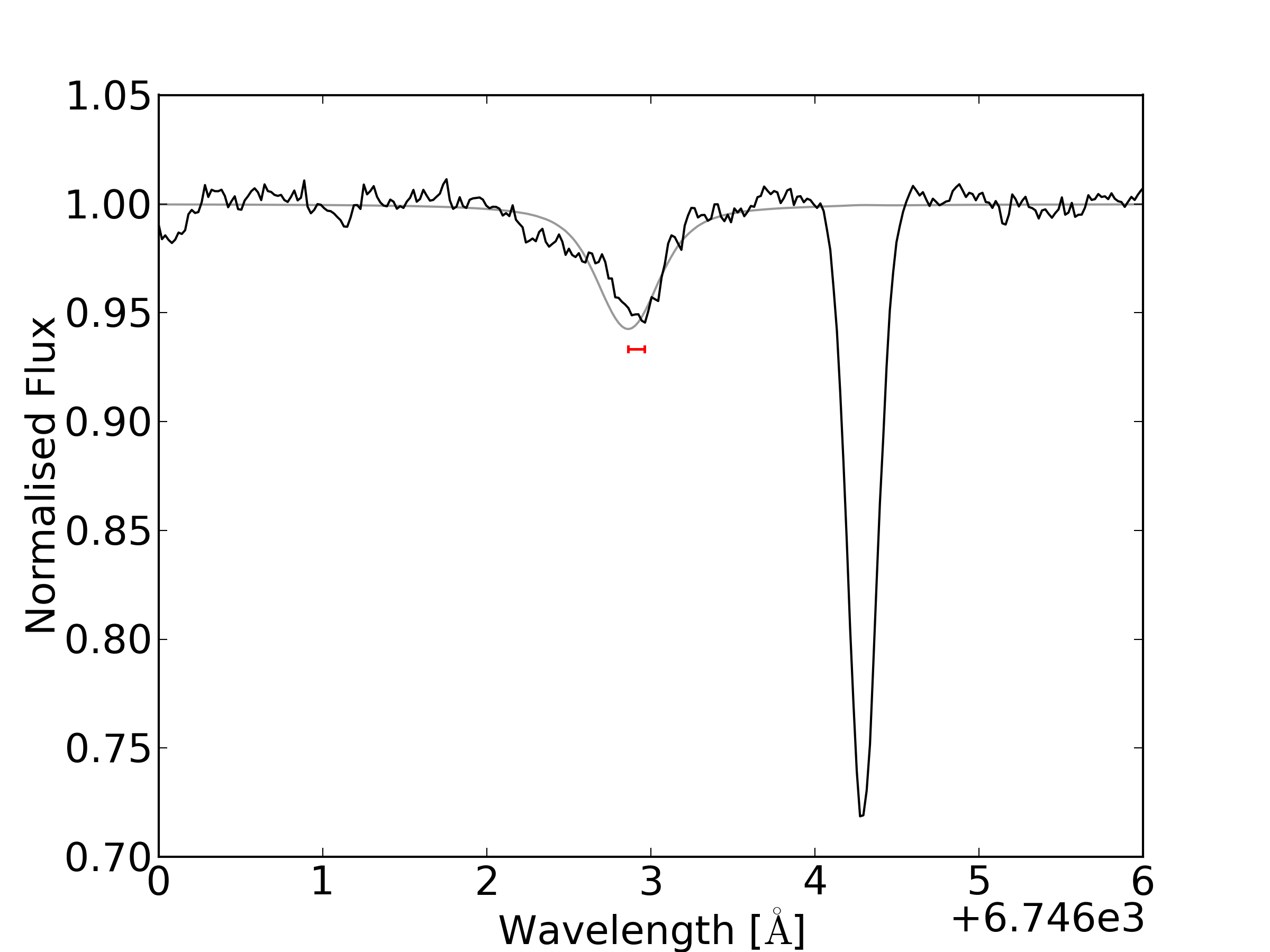}
\includegraphics[scale=0.29]{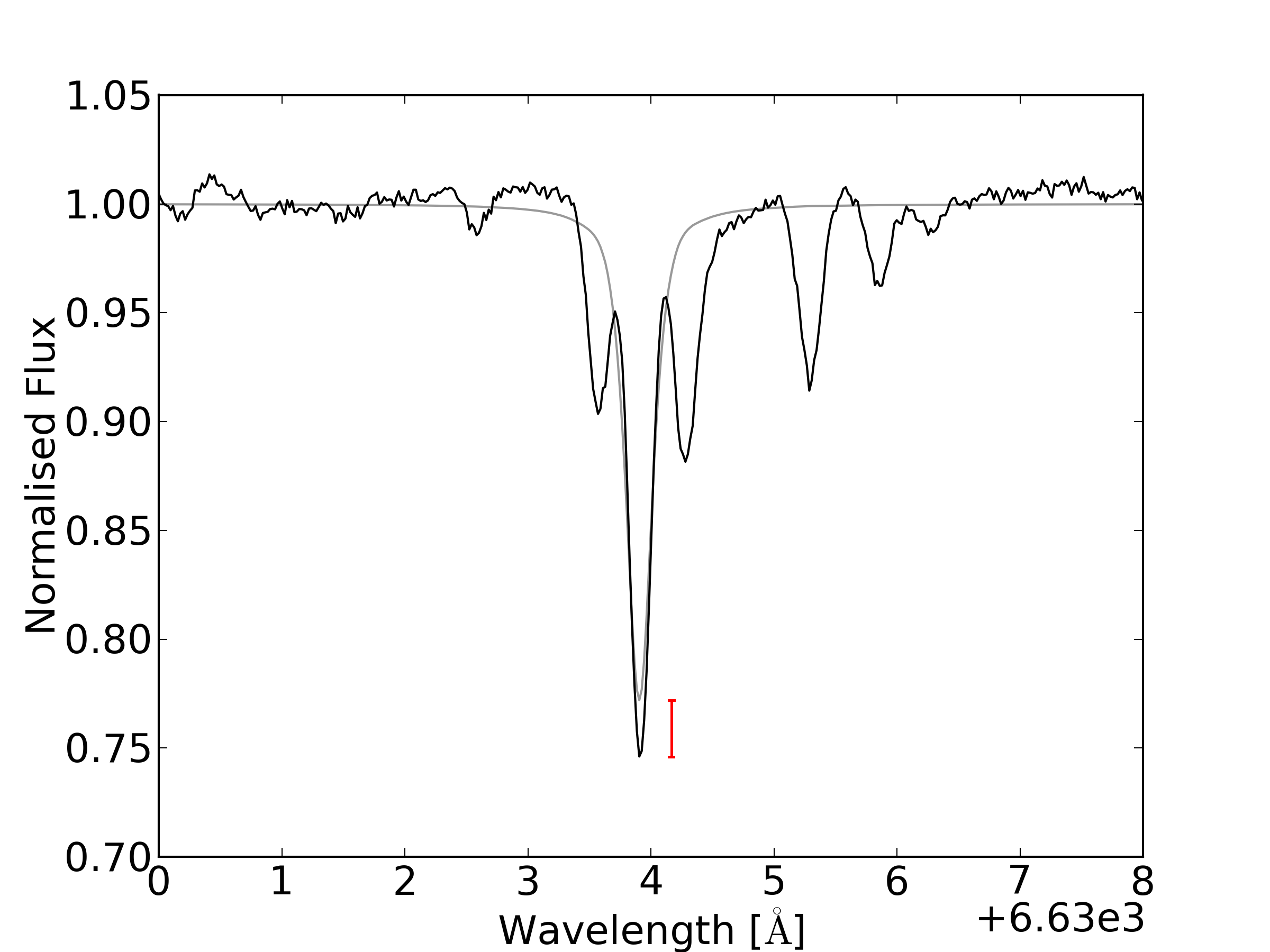}}
\makebox[\textwidth]{
\includegraphics[scale=0.29]{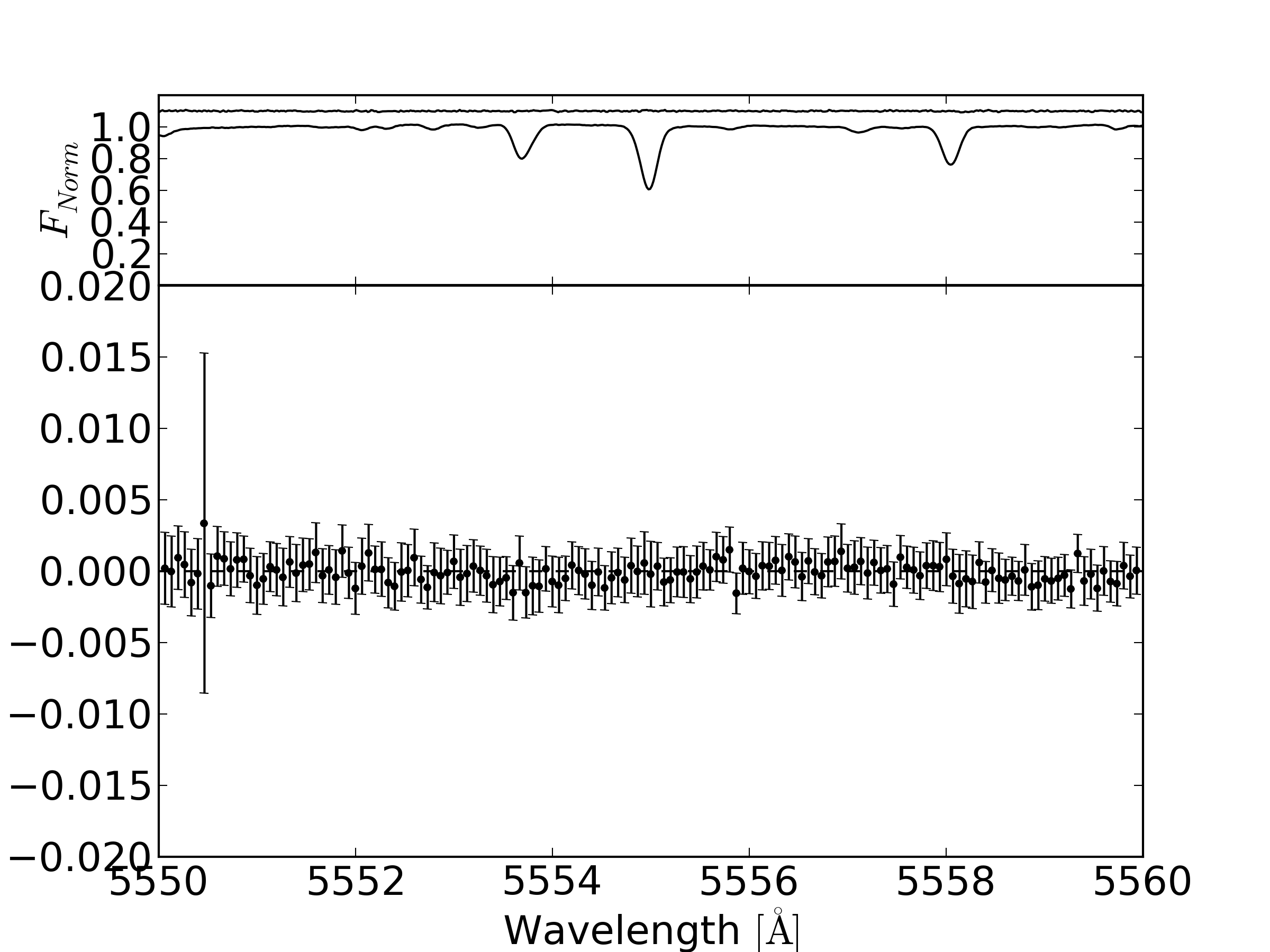}
\includegraphics[scale=0.29]{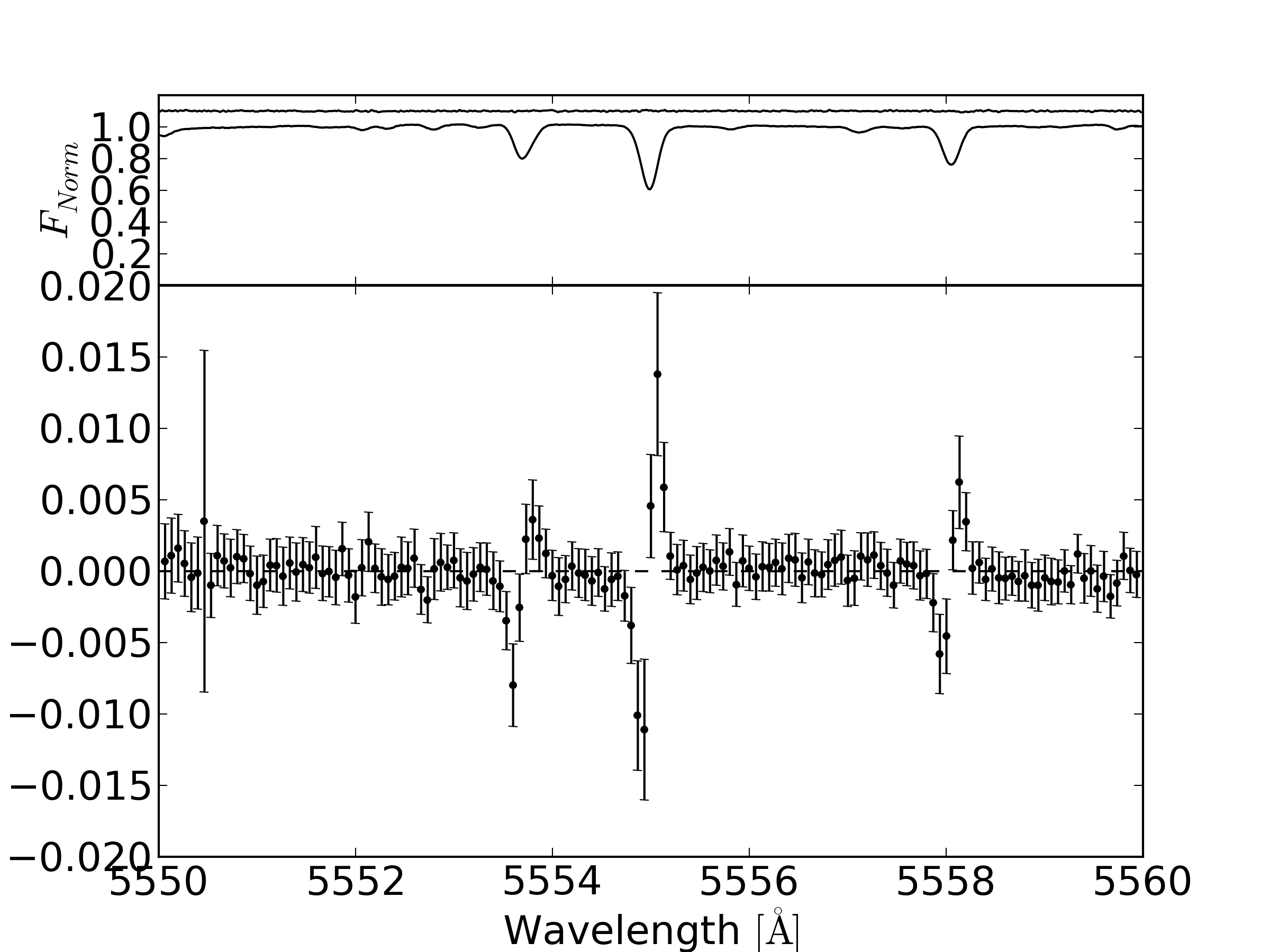}
\includegraphics[scale=0.07]{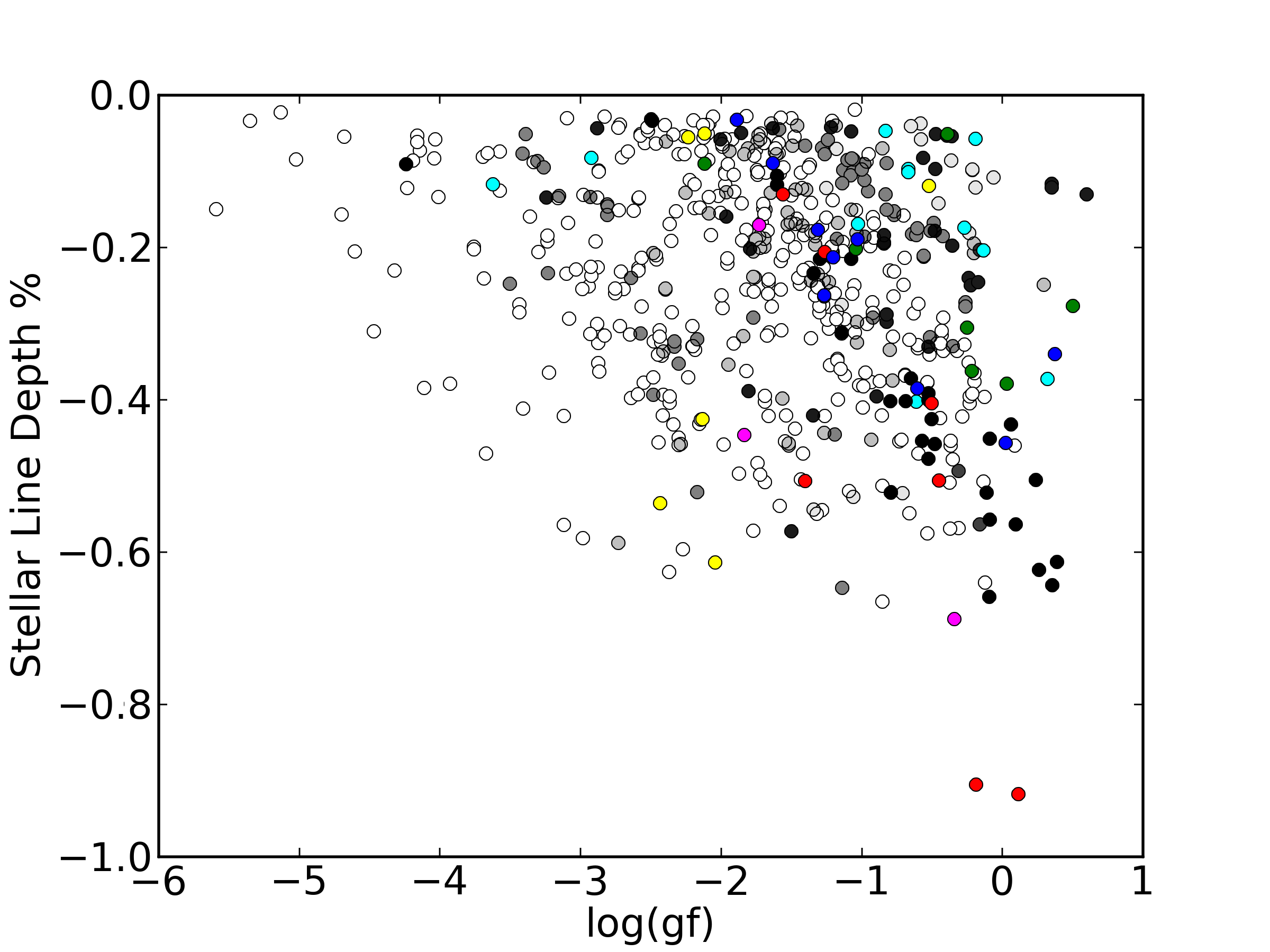}}
\caption{\emph{Top row:} Cases of transitions rejected by the selection criteria. \emph{From left to right:} A crowded zone: Black and blue points are pixels considered as continuum by the median filter and the median value, respectively. The sorted residuals are also plotted, where the 70$\%$ limit (black to red) is over $0.009$. An asymmetrical blend: Red segments represents the wavelength difference. A symmetrical blend: Red segment shows the flux difference. \emph{Bottom row:} The transmission spectrum without \emph{(left)} and with \emph{(middle)} the local shift correction. The stellar spectrum is shown at the top. \emph{Right:} Stellar line depths are generally stronger as greater oscillator strength values. Several species are plotted with different colours.}
\label{fig:selcrit}
\end{figure*}

A Doppler correction for each Echelle order was applied in order to locate observable transitions from the database. We used 4 to 8 transitions to obtain the linear dependence between the rest and observed wavelength difference as a function of the observed wavelength. This procedure was less accurate than calculate the radial velocities by the correlation-cross-function but precise enough to find the Doppler-shifted transitions from the database.

We do not follow the method used in W2004, N2005 and S2008 to correct the variations in the systematic blaze function. Instead, we normalized locally\footnote{5 {\AA} and 7 {\AA} around each transition, equivalent to 250-350 times the spectral resolution, respectively.} around each analyzed transition by implementing a median filter, sigma clipping, and using a third order polynomial (details in Fig.~\ref{fig:norm}). Then, we applied the following automated selection criteria to each spectral line in order to identify those with the best conditions for the further analysis.

\begin{itemize}
\renewcommand{\labelenumi}{\roman{enumi}}
\item Since faint transitions are more difficult to detect, we selected those stellar lines having a minimum line depth of 10\% in the database (i.e. at the solar spectrum).
\item Zones with bad automatic normalization were discarded by obtain pixel by pixel the residuals from the normalized continuum and its median. These values were sorted. Selected transitions have 70\% of sorted residuals below the empirical value of 0.009.
\item To discard blended lines we calculated the difference in wavelength and normalized flux between the pixel with the lowest value (line core) and the center of a Voigtian fit. Such difference for selected lines were less than the empirical value of 0.06{\AA} ($\sim$3 pixels) in wavelength and less than 0.06 in normalized flux. 
\item Stronger transitions are easier to detect in the transmission spectrum. Therefore, we use the oscillator strength as a proxy. We selected those transitions having an oscillator strength value greater than the value of Na D ($log(g_if_{ij})=-0.184$) which was detected in HD 209458b using this data set (S2008). The transitions strength of Ca, Fe, Mg, Mn, Na and Sc indeed empirically correlate with oscillator strength.
\end{itemize}

Finally, we note that most of the lines present an ``s'' shape in the transmission spectrum.  \citet{Jensen2012_Halpha} report the same ``s'' shape for H$\alpha$ and gives some possible interpretations. However, in our analysis we found that this unusual feature is not planetary phase-correlated because it also appears when comparing only out-of-transit data. To correct these relative shifts between spectra we calculate the centre of each stellar line by the Voigtian fit. This centre was used to obtain the difference in wavelength between the absorption in a reference exposure and the rest of frames. Then, we align the spectral line for each exposure to a common centre using a cubic spline re-sampling. Figure ~\ref{fig:selcrit} summarizes the selection criteria by examples.

\subsection{Telluric corrections}
\label{sec:telluric_correction}

Spectra recorded using ground-based instruments suffers from the imprints of Earth's atmosphere which could vary in time due to different reasons. Therefore, those data must be corrected from telluric features producing spurious signals in the transmission spectrum. A powerful method to recognise telluric lines was implemented using the airmass in the radiative transfer equation. The solution of this equation assuming only absorption is $ln \left( I/I_0 \right) = Nk_\lambda s$; where $I$ is the intensity at certain wavelength $\lambda$, $I_0$ is the source intensity, $Nk_\lambda$ is the optical depth at zenith and \emph{s} is the airmass. We ignore slit-losses as first trial.

\begin{figure}[t]
\centering
\includegraphics[scale=0.4]{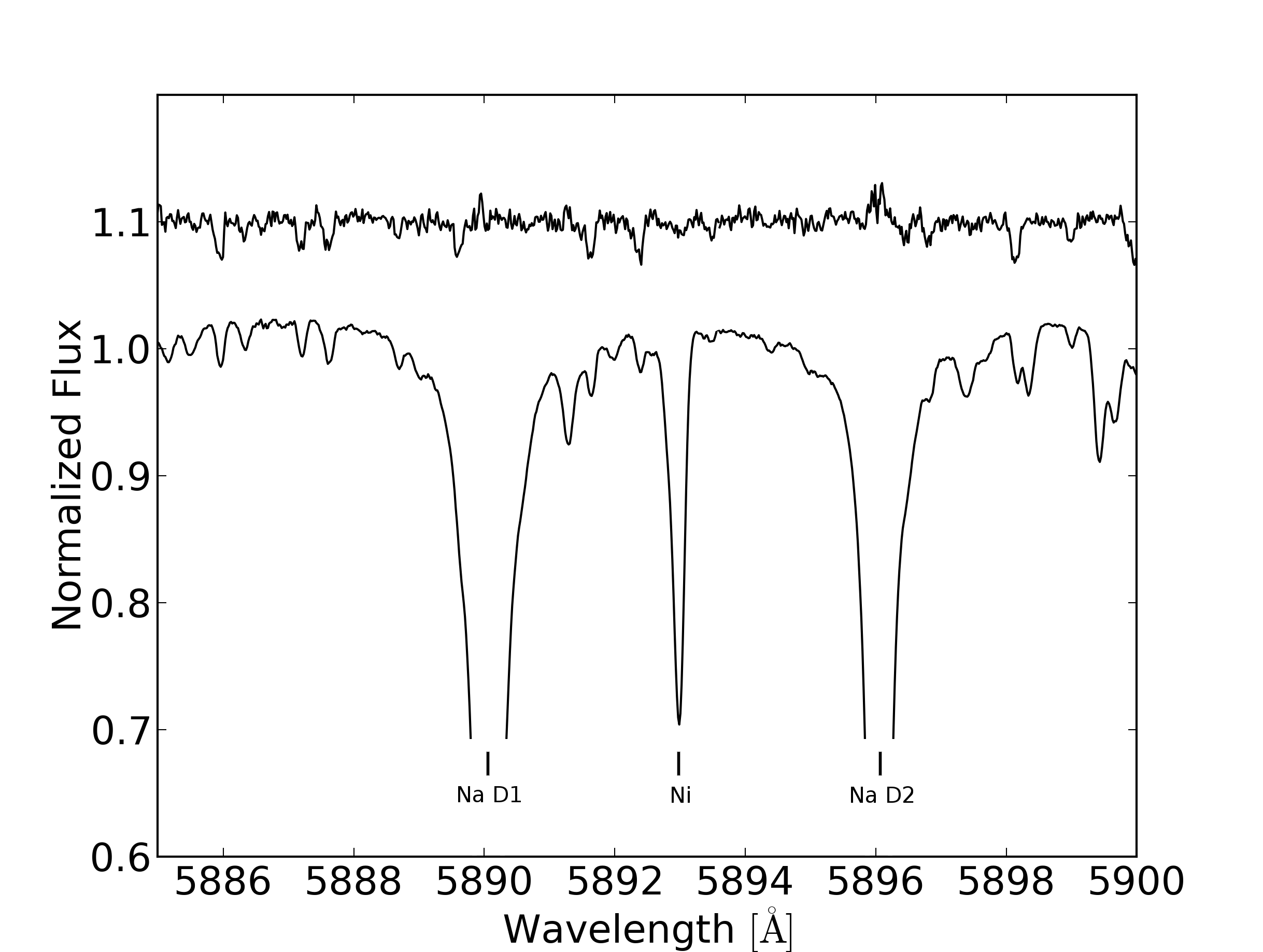}
\caption{Example of telluric absorptions in HDS spectra around the Na I D. HD 209458 normalized spectrum is shown while the telluric spectrum is shown above the latter with an offset of 0.1.}
\label{fig:telluric_example}
\end{figure}

We derived $Nk_\lambda$ and its error through a linear regression to $ln(I)$ as a function of the airmass $s$ at a mid-exposure. This procedure was done only using out-of-transit spectra because while the transit starts a relatively abrupt variation in $F_\lambda$ is expected if a detectable element is present in the exoplanetary atmosphere blocking photons at $\lambda$. This method yields a telluric spectrum (e.g.\ Fig.~\ref{fig:telluric_example} top) which is clearly consistent with the telluric spectrum shown in Figures 1 from L2009 or S2008. Any slit-losses were thus empirically shown of second order.

Each spectrum was divided by $e^{Nk_{\lambda}(s_i-s_{Ref})}$, where $s_{Ref}$ is the average airmass of in-transit spectra. We note that some strong tellurics are not fully corrected as features appears in telluric lines positions in the transmission spectrum. We attribute this to short timescale variations in telluric species content or second-order deviations from the linear dependence $ln(I) \propto s$, like slit-losses. We paid special attention when a feature is present in both the transmission and the telluric spectrum, like in Na D vicinity.

\subsection{Features in the transmission spectrum}
\label{sec:features_transmission_spectrum}

For each transition, we created a central (C) pass-band centred in the interest line and two adjacent - blue (B), red (R) - control pass-bands. From R2008, and \citet{Jensen2011_alkali_survey, Jensen2012_Halpha} we note that features in the transmission spectrum falls within stellar lines profiles because even if spectral lines are broader in atmospheres such as HD 209458b the strongest absorption occurs at the core. This fact suggested to select pass-band sizes proportional to the stellar line width. Pass-band sizes were given by $\times3$, $\times4.5$ and $\times6$ the width of the Voigtian fit to each stellar transition. We note that originally C2002 used three central band-passes because they did not know the precise width of the sought feature.

To automatically detect features in the transmission spectrum we calculate the relative flux based both with R2008 and S2008 methods. The depth of the light curve should be consistent in both cases.

\begin{figure}[t]
\centering
\includegraphics[scale=0.4]{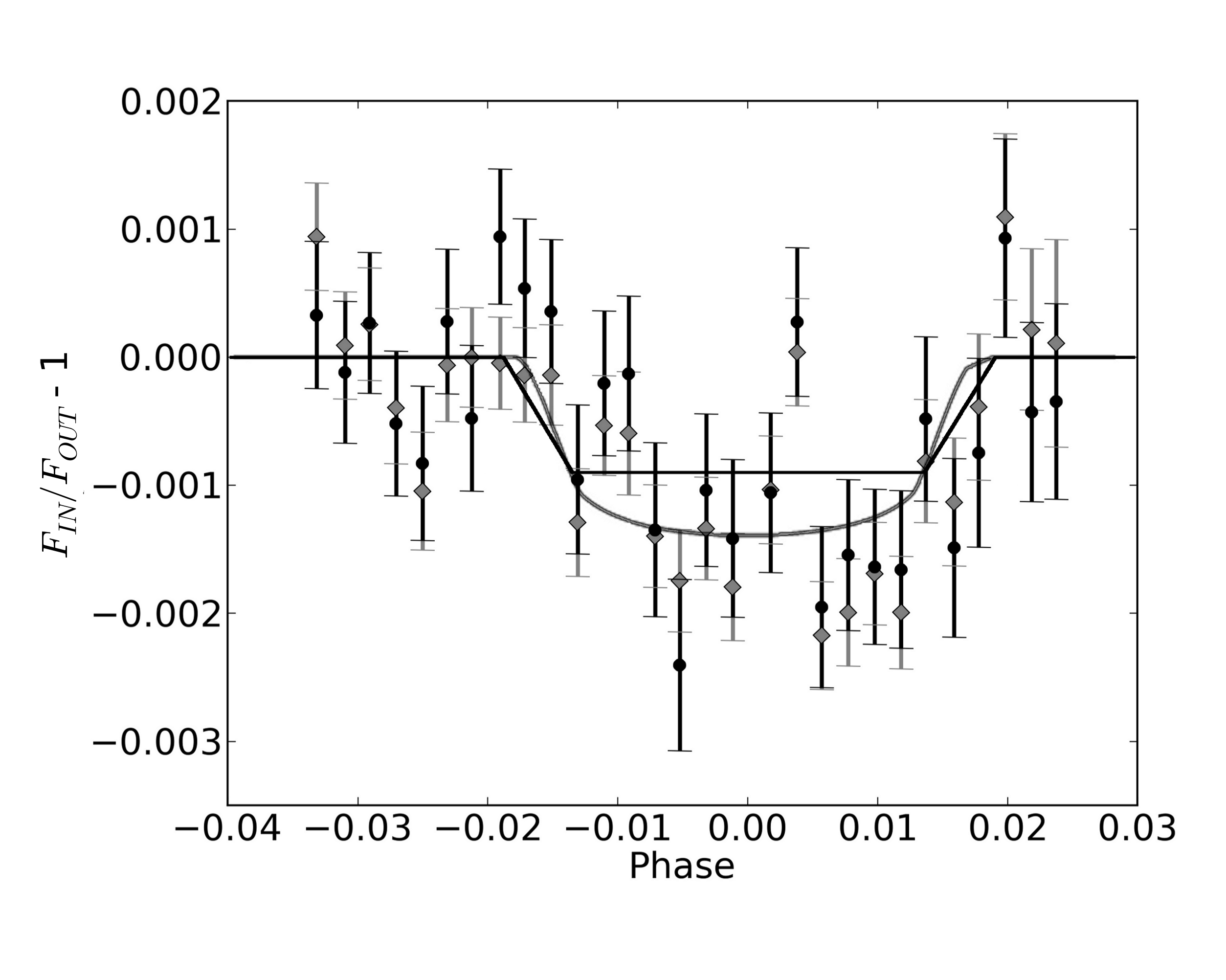}
\includegraphics[scale=0.4]{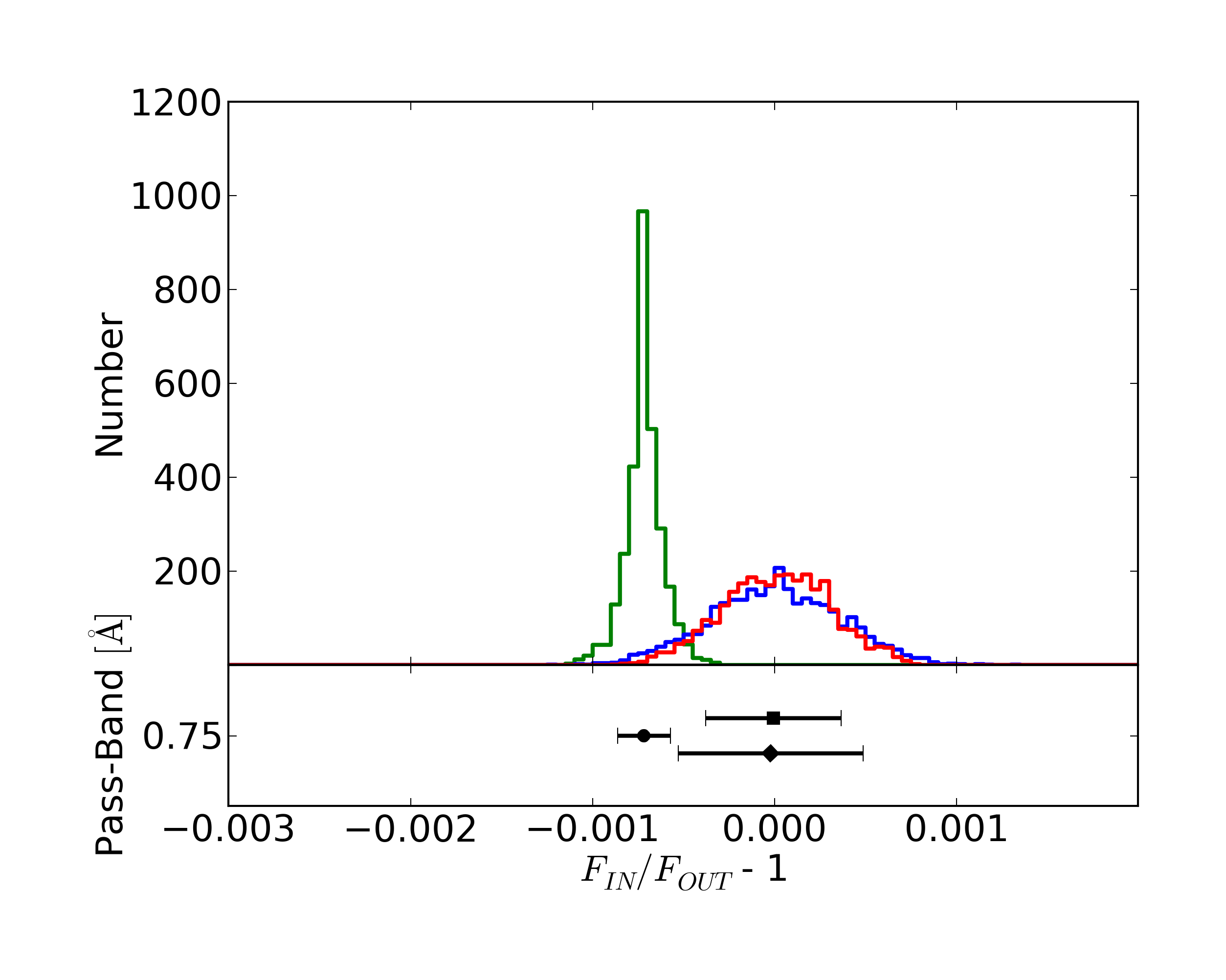}
\caption{Comparison from Na I D with previous detection reported in this data set where the used pass-band was 0.75 {\AA}. \emph{Top panel:} Transit photometry derived by S2008 (grey) and this work (black). The correction proposed by S2008 for CCD non-linearity was applied in this work. Solid lines represent the transit fits for this work (black) and S2008 (grey). The latter used a scaled version of HST photometry integrated over 582 and 638 $nm$ which includes limb darkening contribution. \emph{Bottom panel:} R2008 based analysis for the Na I D transmitted signal and its uncertainty derived from the bootstrap analysis. Green colour represent the \emph{in-out} scenario, blue colour \emph{in-in} scenario, and red colour the \emph{out-out} scenario. The centre and 1$\sigma$ of each distribution is shown at the bottom.}
\label{fig:s2008_comparison}
\end{figure}

\subsubsection{R2008 based analysis.}
\label{sec:R2008}

This analysis visualises the transmission spectrum along wavelength. In-transit ${\cal F}_{In}$ and out-of-transit ${\cal F}_{Out}$ spectra were stacked separately reaching a $SNR \sim 1600$ and $SNR \sim 1200$, respectively. ${\cal F}_{In} \left( \lambda  \right)$ and ${\cal F}_{Out} \left( \lambda  \right) $ errors were obtained with the standard deviation of normalized flux by pixel across in-transit and out-of-transit frames. The relative flux $F_{Rel}\left( \lambda \right)= \overline{{\cal F}_{In}} / \overline{{\cal F}_{Out}}$ was computed and then the transmitted signal $F_{Trans}$ was calculated for each pass-band as follows

\begin{displaymath}
F_{Rel}(C) = \frac{\overline{{\cal F}_{In}(C)}}{\overline{{\cal F}_{Out}(C)}} \, , \;
F_{Rel}(B) = \frac{\overline{{\cal F}_{In}(B)}}{\overline{{\cal F}_{Out}(B)}}  \, , \;
F_{Rel}(R) = \frac{\overline{{\cal F}_{In}(R)}}{\overline{{\cal F}_{Out}(R)}}
\end{displaymath}
\begin{equation}
F_{Trans} = F_{Rel}(C)  - \left[ F_{Rel}(B) + F_{Rel}(R) \right] / 2
\end{equation}

\bigskip

To deal with systematic and random errors a statistical diagnostic of the stability of a measured absorption was done by bootstrapping analysis.

Three different scenarios were created following R2008's Section 3.2. (1) An \emph{out-out} comparison where a randomly selected subset of \emph{out-of-transit} data (7 frames) were used like in-transit data and the non-selected \emph{out-of-transit} (4 frames) were used like out-of-transit exposures, each data subset was then used to calculate the transmitted signal. Since there should be no difference among the transmitted signal for these population, the resulting distribution is expected to be centred at zero. (2) Similarly, an \emph{in-in} comparison where a randomly selected subset of \emph{in-transit} exposures (11 frames) were compared to the rest of \emph{in-transit} data. The resulting distribution is also expected to be centred at zero. (3) An \emph{in-out} comparison where an increasing number of spectra (up to half of the sample) were removed from the \emph{in-transit} sample and the remaining were compared with the \emph{out-of-transit} data set. The resulting distribution is expected to be centred at the measured value of the transmitted signal using all data set.

The number of iterations was selected 3000 because approximately over this number the width of distributions in the \emph{in-in, out-out} and \emph{in-out} scenarios do not increase significantly. We considered 1$\sigma$ of the \emph{in-out} scenario as the uncertainties for our later results.

\subsubsection{S2008-based analysis.}
\label{sec:S2008}
This analysis visualises the light curve derived from the transmission spectrum in a specific wavelength. We perform the spectrophotometry deriving $F_{Rel}(t)$ for each exposures using the same pass-bands described in Sect.~\ref{sec:features_transmission_spectrum}. Derived light curves were normalized and the transmitted signal $F_{Trans}$ (the light curve depth) was obtained by

\begin{equation}
F_{Rel}(t) = \frac{2\overline{F_{C}}}{\overline{F_{B}}+\overline{F_{R}}}\, , \; F_{Trans} = \overline{F_{Rel} \left( t_{In} \right)} - \overline{F_{Rel} \left( t_{Out} \right)}
\end{equation}

where the uncertainty of the relative flux $\sigma_{F_{Rel}(t)}$ was estimated by considering only photon noise.

Contrary to S2008, however, we do not fit a model with stellar limb-darkening. Instead, we measure the absorption as the difference between the in-transit and out-of-transit averages. Stellar limb darkening is not expected to significantly affect the transmitted signal since we used narrow band-passes and spectra were normalized by the local continuum. As demonstrated by R2008 using stellar models, and depending on the spectral line, the error for not considering limb-darkening accounts to a negligible $\lesssim 10^{-5}$. On the other hand, using the wrong limb-darkening law can be misleading (see Sect.~\ref{sec:NaI}).

\begin{figure}[!ht]
\centering
\includegraphics[scale=0.4]{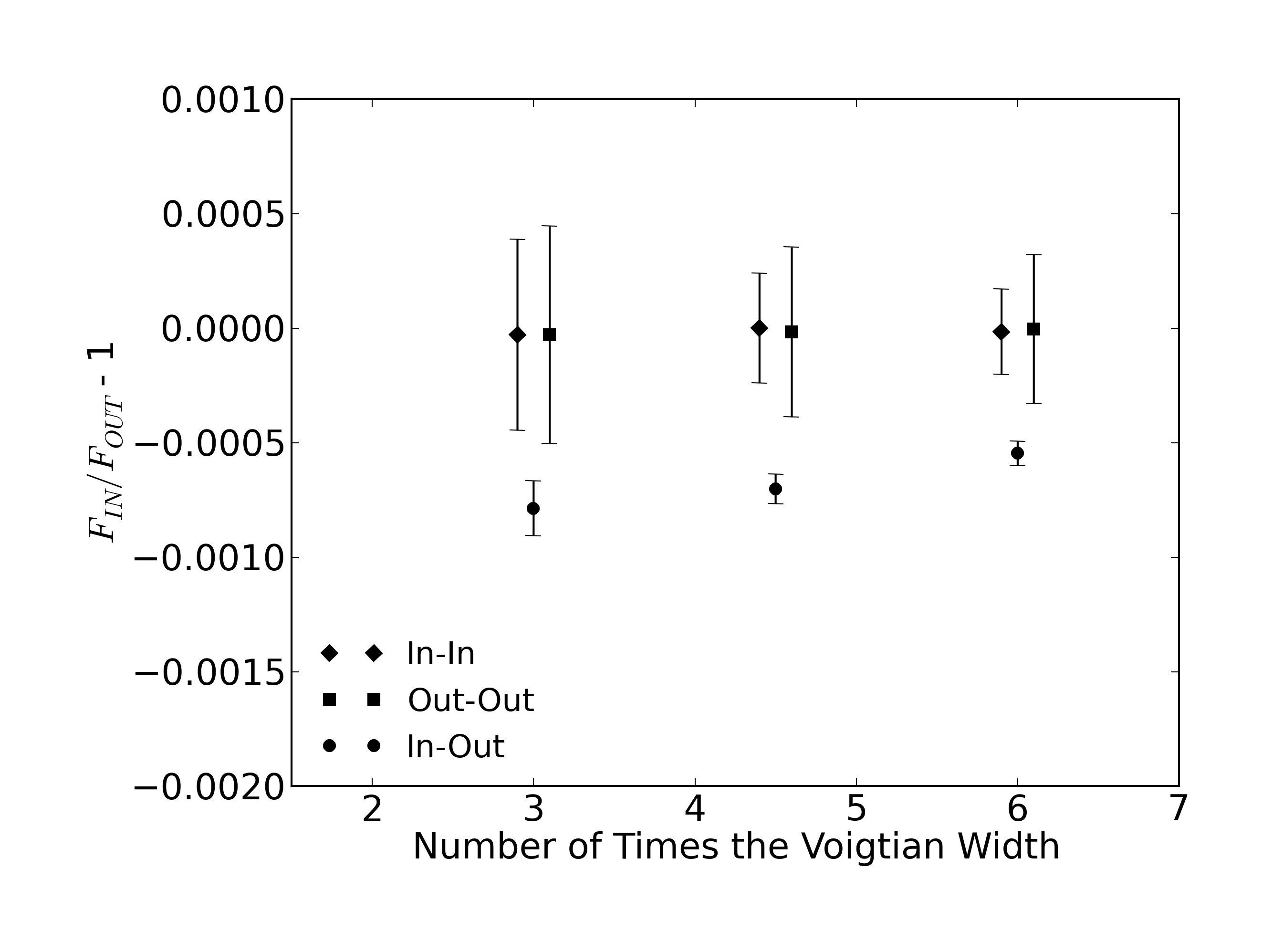}
\includegraphics[scale=0.4]{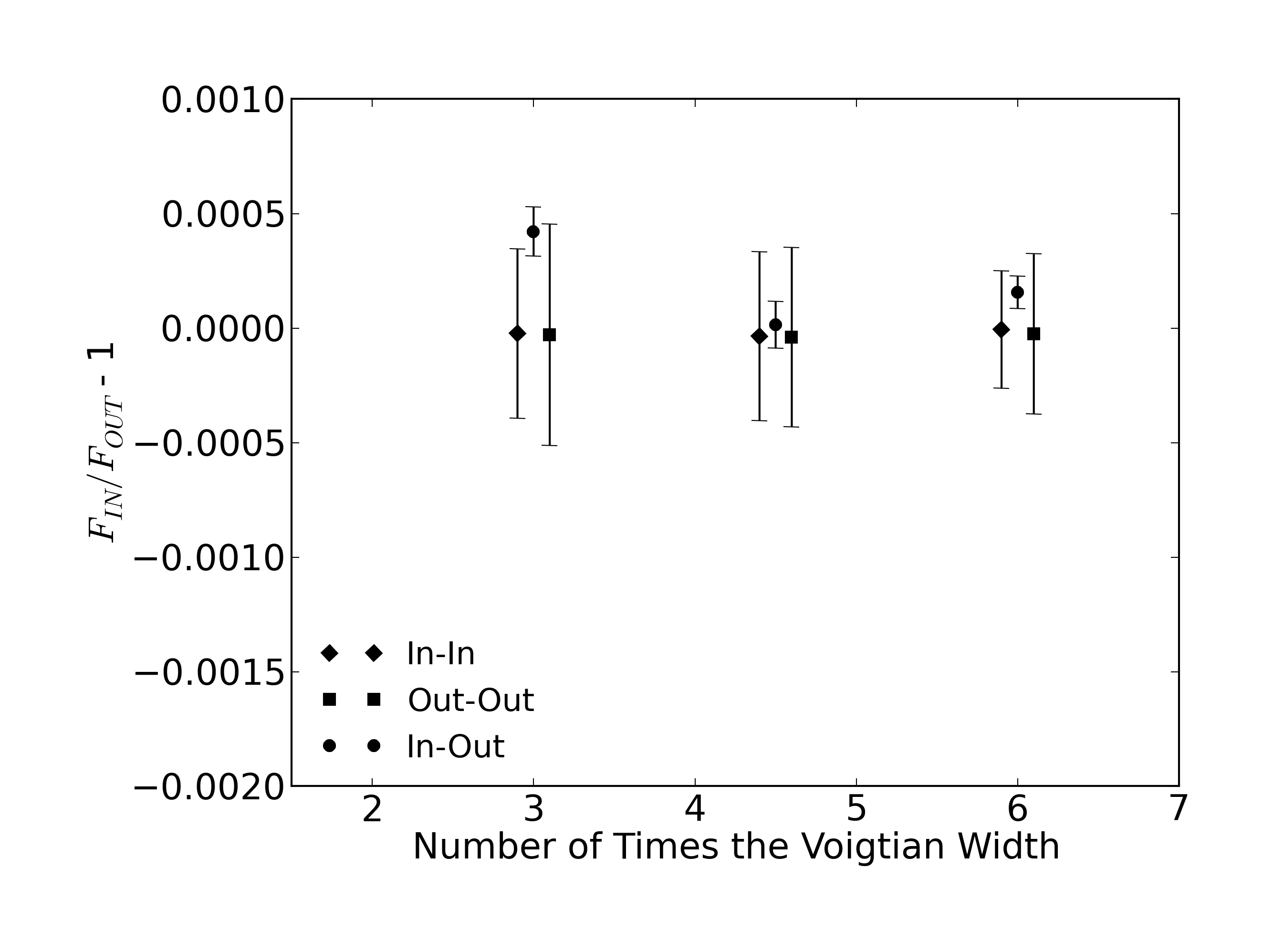}
\includegraphics[scale=0.4]{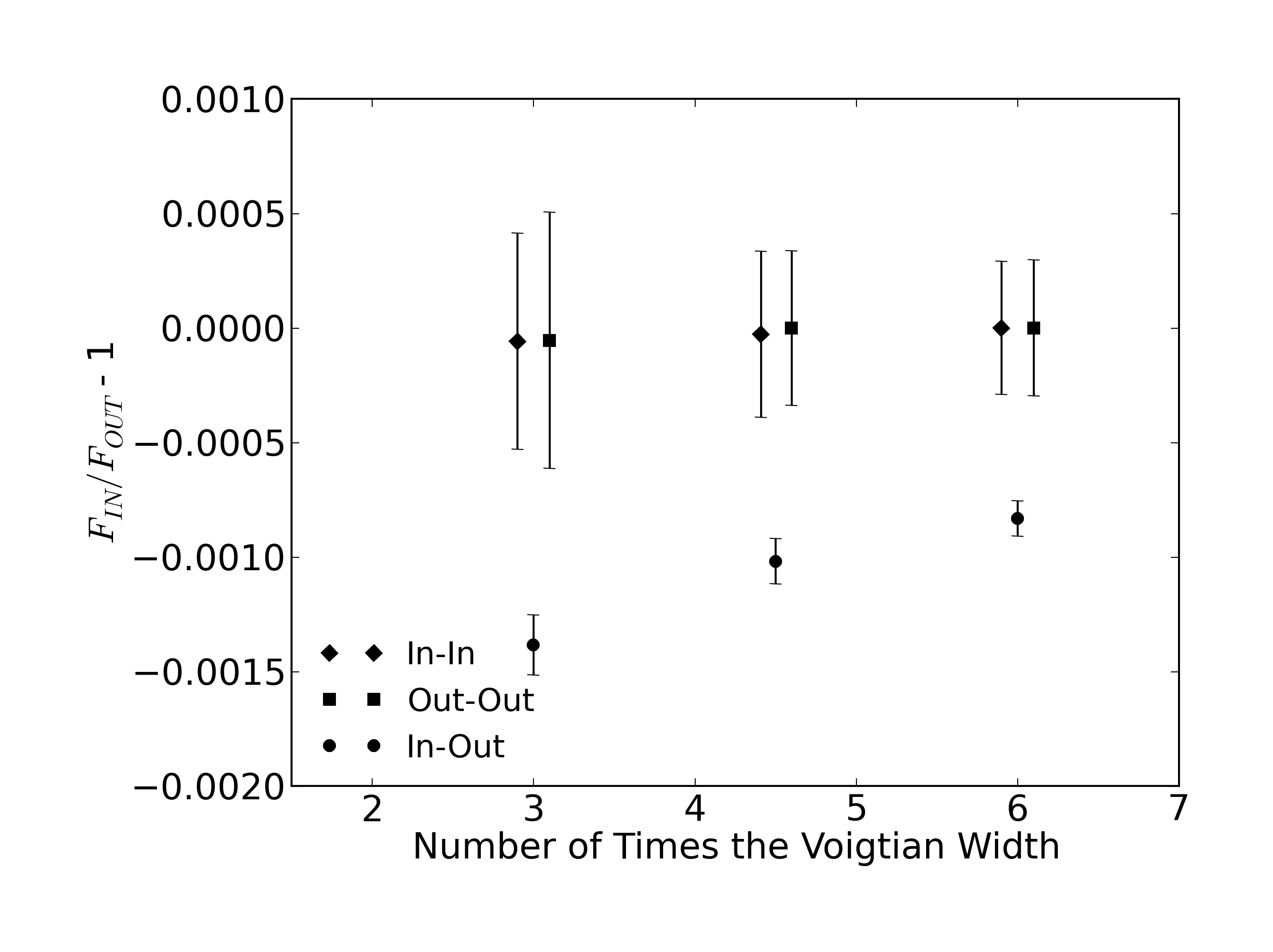}
\caption{Transmitted signals of the Ca I transitions at 6162.27 {\AA} (top panel), 6439.1 {\AA} (middle panel), and 6493.78 {\AA} (bottom panel) using pass-bands of $\times$3, $\times$4.5, and $\times$6 the stellar line width derived from fitting a Voigt profile. The significance of measurements were derived from the \emph{in-out} scenario in the bootstrap analysis described in Sect.~\ref{sec:R2008}. The noise levels were derived from the \emph{in-in} and \emph{out-out} scenarios of the bootstrap analysis.}
\label{fig:Ca_trans_bands}
\end{figure}

\begin{figure}[t]
\centering
\includegraphics[scale=0.4]{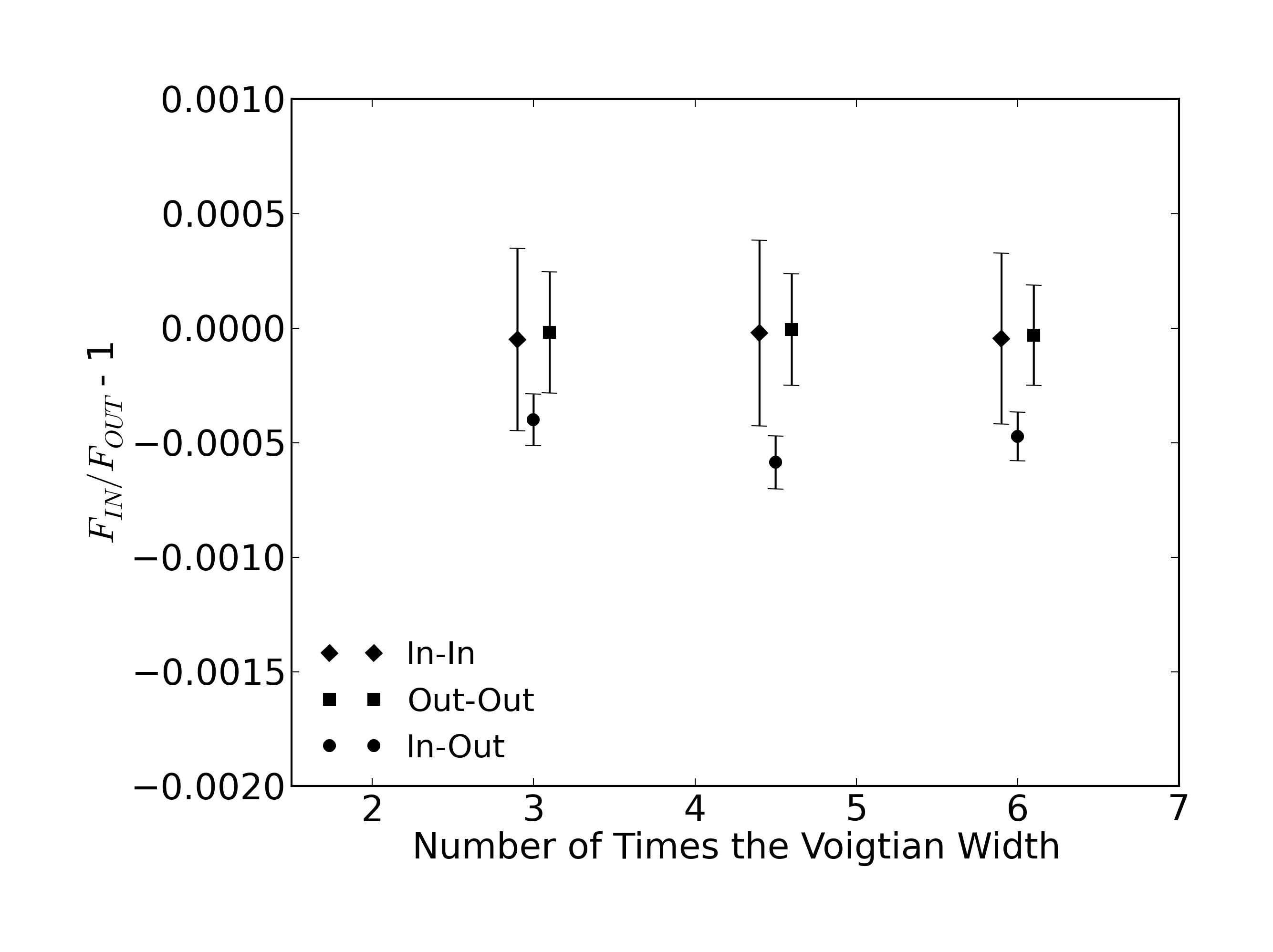}
\caption{Same as Fig.~\ref{fig:Ca_trans_bands} but for Sc II at 5526.79 {\AA}.}
\label{fig:Sc_trans_bands}
\end{figure}

\section{Results and analysis}

Previous Sodium and Hydrogen detections were confirmed and new Calcium and possibly Scandium features resulted in significant absorption in the transmission spectrum from 29 transitions satisfying our selection criteria described in Sect.~\ref{sec:data_preparation} and over 2200 transitions automatically analysed in the twenty-nine spectra.

\subsection{Na I}
\label{sec:NaI}

As it is expected from the previously analysed data set (S2008), absorption excess in Na I D was measured at a level of $-0.071\pm0.016\%$ ($\times$3 band), $-0.043\pm0.012\%$ ($\times$4.5 band) and $-0.045\pm0.010\%$ ($\times$6 band). The averaged Voigtian width in Na D transitions was 0.21 {\AA}. We note that the effect of stellar limb darkening is an order of magnitude less than our obtained statistical error.

As comparison, C2002, S2008 and L2009 measured an absorption of $-0.023\pm0.006\%$, $-0.135\pm0.017\%$ and $-0.109\pm0.026\%$ in the Sodium doublet using their own bandwidths, respectively. Selecting the same pass-bands used in S2008, our detection level was $-0.072\pm0.015\%$ (0.75 {\AA}). Fig.~\ref{fig:s2008_comparison} top panel's shows the light curve derived for Na I D absorption excess using the 0.75 {\AA} pass-band while bottom panel shows the associated bootstrap distributions (top) where we specify its central values and 1$\sigma$ uncertainties (bottom). We note that almost all our light curve points are within S2008's 1$\sigma$ values. In fact, using our same transit-fitting modeling on S2008's data (average-in minus average-out) we obtain a difference between their depth and ours of $0.016\%$, which is in agreement with the errors. The disagreement with their published Na D absorption ($-0.135\pm0.017\%$) therefore seems to come primarily from their choice of fitting a model with a limb darkening law adapted from a broadband observation.

\begin{figure*}[t]
\centering
\includegraphics[scale=0.35]{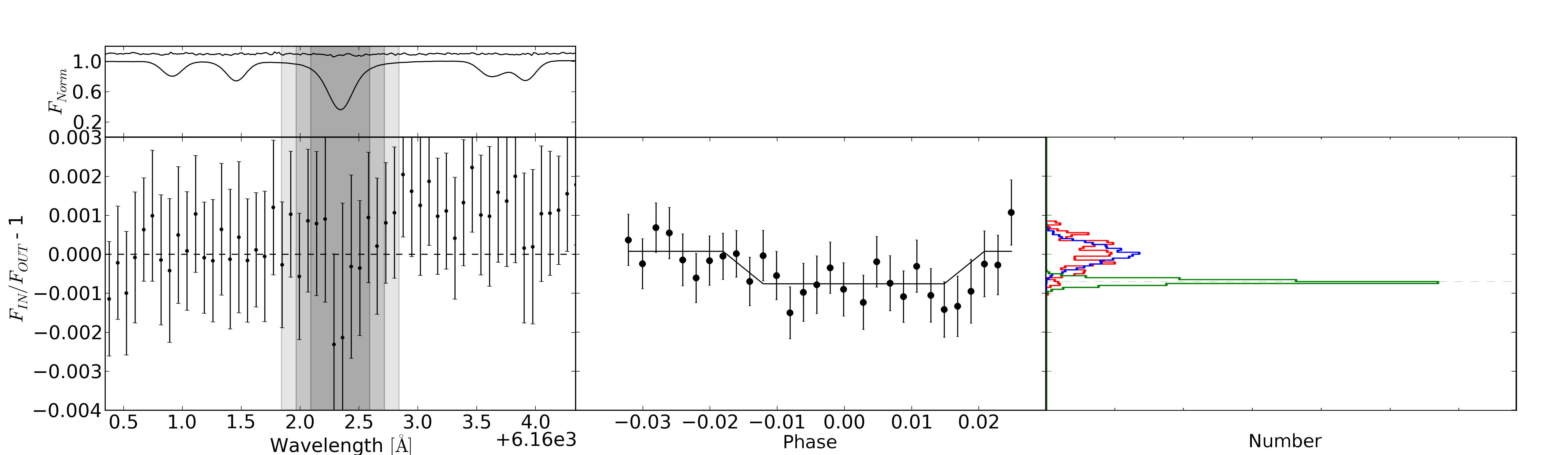}
\includegraphics[scale=0.35]{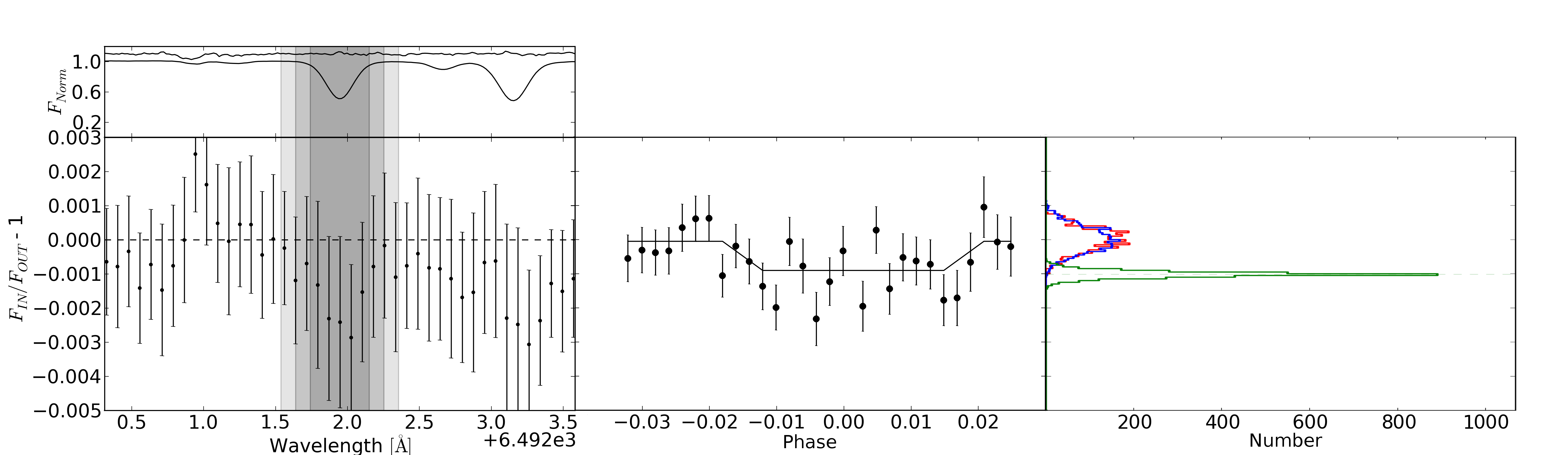}
\includegraphics[scale=0.35]{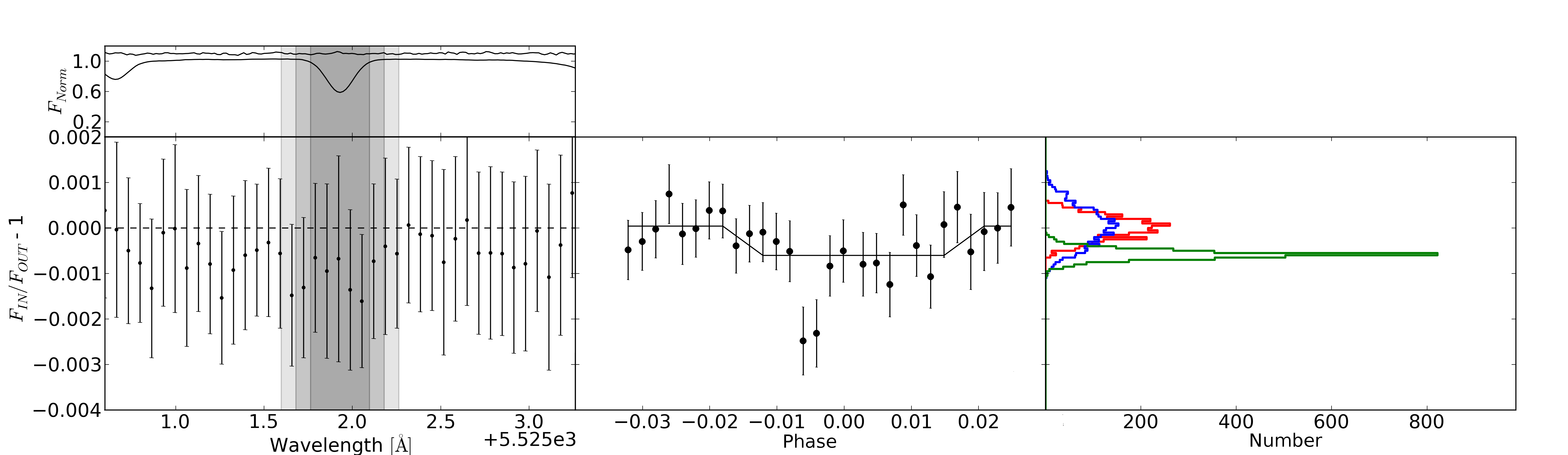}
\caption{Analysis of the transitions of Ca I at 6162.27 {\AA}, Ca I at 6493.8 {\AA}. and Sc II at 5526.79 {\AA} are shown in the first, second and third row, respectively. \emph{Left panels:} Absorption excess in the binned transmission spectrum. For reference, the normalised stellar spectrum and the telluric spectrum (0.1 offset and 3 times magnified) are shown at the top of each panel. The shaded zones represent the pass-bands sizes. \emph{Middle panels:} Light curves derived from spectrophotometry using the $\times$4.5 band both Ca I transitions and for the Sc II. Flat light curves represents the averages of in-transit and out-of-transit data. \emph{Right panels:} Derived bootstrap distributions showing the detection stability. the colours represents: blue is the \emph{in-in} distribution, red is the \emph{out-out} distribution, and green is the \emph{in-out} distribution.} 
\label{fig:analysis_figures}
\end{figure*}

\subsection{Ca I}

Ca I transitions at 6162.17, 6439.08 and 6493.78 {\AA} passed all four selection criteria over eighty-five Ca transitions analyzed. Transmitted signals for each Ca I transition and selected pass-bands are shown in figure~\ref{fig:Ca_trans_bands}.

Ca I at 6162.27 {\AA} has an absorption excess detected at $10 \sigma$, the measurements are $-0.079\pm0.012\%$ ($\times$3 band), $-0.070\pm0.007\%$ ($\times$4.5 band) and $-0.055\pm0.005\%$ ($\times$6 band). The Voigtian fit to the stellar line gives a width of 0.17 {\AA}. Measurements through the different pass-band sizes at 6439.1 {\AA} are within $1\sigma$ error derived from the \emph{in-in} and \emph{out-out} scenarios of the bootstrap analysis and therefore is not considered as a detectable signal. These values are $0.042\pm0.010\%$, $0.001\pm0.010\%$ and $0.016\pm0.007\%$ in the $\times$3, $\times$4.5 and $\times$6 bands, respectively. The Voigtian width used for the pass-band sizes was 0.15{\AA}. Finally, the Ca I transition at 6493.78 {\AA} present a strong absorption excess at a level of $-0.138\pm0.013\%$ ($\times$3 band), $-0.102\pm0.010\%$ ($\times$4.5 band) and $-0.083\pm0.008\%$ ($\times$6 band), where the stellar Voigtian width was 0.14 {\AA}.

We derived a radial extent\footnote{$R_\lambda / R_p = \left[ 1+ \left( 1 - R_\star^2 / R_p^2 \right) \left( F_{In} / F_{Out} -1 \right)_\lambda \right]^{1 / 2}$} of $2537\pm391$km (4.6$\times$H), $2267\pm212$km (4.1$\times$H), $1770\pm171$km (3.2$\times$H) for Ca I at 6162.27 {\AA}; $4419\pm429$km (8.0$\times$H), $3271\pm325$km (5.9$\times$H), $2676\pm252$km (4.9$\times$H) for Ca I at 6493.78 {\AA} in $\times$3, $\times$4.5, and $\times$6 pass-bands, respectively. Where $H$ is the scale height and we used $R_p=1.38R_{Jup}$ \footnote{$R_{Jup}$=71492 [km]} and $R_\star/R_p = 0.11384/0.01389$ \citep{Southworth2010_transiting_EP_studies}.

Transitions at 6162.27 {\AA} and 6493.78 {\AA} shows the expected rise of absorption toward smaller pass-bands. Surprisingly, Ca I at 6439.1 {\AA} presents no significant signal. However, we believe that this could be due to some extra systematics from that particular region of the detector, which is near the CCD edge and receives about $20\%$ less photons ($SNR\sim250$ per pixel at the middle of the line profile). This is further supported by the fact that not even a hint of the tellurics in the region are detected even though they have a $log(g_if_{ij})$ as strong as that of tellurics positively detected in different regions.

\subsection{Sc II}

Thirty-six Sc transitions are present in the considered optical range, only the Sc II transition at 5526.79 {\AA} passed our selection criteria. Figure~\ref{fig:Sc_trans_bands} shows results computed for each selected pass-band and their significance from the bootstrapping.

The detection is at $>4\sigma$, at a level of $-0.059\pm0.012\%$ and $-0.047\pm0.011\%$ for the $\times$4.5 and $\times$6 pass-bands, respectively. The Voigtian width of the stellar line used in the pass-band sizes is 0.11 {\AA}.

However, for the $\times$3 band the measured value is within $1\sigma$ error derived from the \emph{in-in} and \emph{out-out} scenarios of the bootstrap analysis. Hence, this $-0.040\pm0.012\%$ measurement can not be considered a detectable signal. As no telluric transition is reported in the used database, possible explanations are systematic errors or the presence of this elements in the exoplanetary atmosphere producing an unexpected emission, which may produce a fainting effect in the rise of absorption toward smaller pass-bands.

Nevertheless, the light curve in the $\times$4.5 pass-band clearly shows the transit shape and is supported by the bootstrap analysis. Assuming that the absorption comes from the exoplanetary atmosphere and using HD 209458 physical parameters we obtained a radial extent of $1896 \pm 379$km (3.4$\times$H), and $1532 \pm 349$km (2.8$\times$H) using the $\times$4.5 and $\times$6 pass-bands.

\subsection{H$\alpha$}

\begin{figure}[t]
\centering
\includegraphics[scale=0.4]{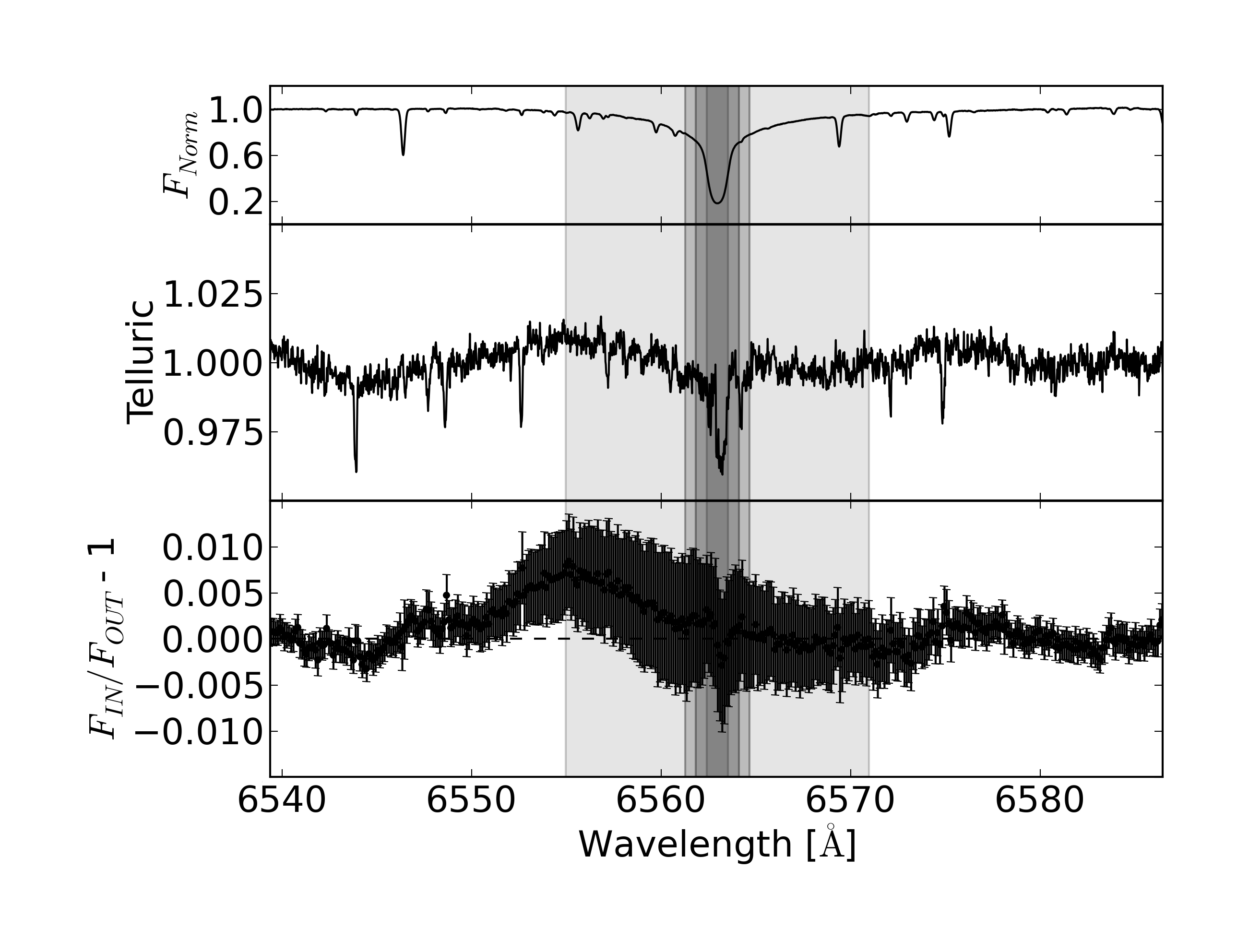}
\includegraphics[scale=0.4]{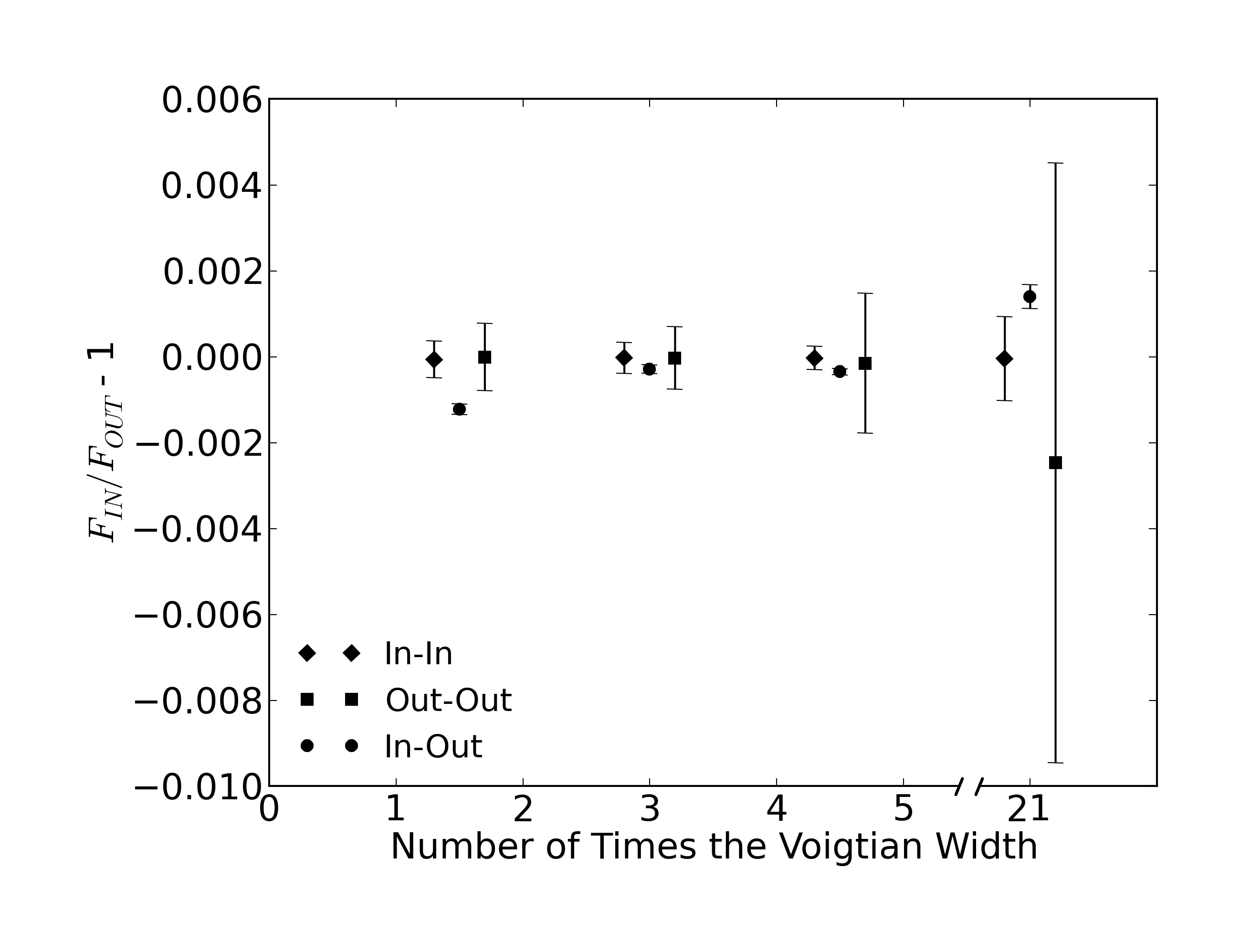}
\caption{\emph{Top panel:} From top to bottom, HD 209458 spectrum near $H\alpha$ transition, our derived telluric spectrum, and the transmission spectrum. The selected pass-bands are $\times$1.5, $\times$3, $\times$4.5 the stellar line width, and the 16 {\AA} pass-band used in \citet{Jensen2012_Halpha}. \emph{Bottom panel:} Transmitted signal measured in each used spectral window, including the $\times$21 (16 {\AA}). This panel shows a possible detection of absorption excess due H$\alpha$ in the narrower pass-band.}
\label{fig:Halpha}
\end{figure}

Recently \citet{Jensen2012_Halpha} report H$\alpha$ detection in the atmosphere of HD 189733b. They analyzed others exoplanets, including HD 209458b. Our algorithm did not automatically analyzed this Hydrogen transition because we used a narrower 5 {\AA} and 7 {\AA} window to normalise locally. So that this transition was rejected by our selection criteria.

Changing the window size to 25 {\AA}, using an estimated FWHM of 1.5 {\AA} for the H$\alpha$ transition and rejecting pixels within 10 times the FWHM (see Sect.\ref{sec:data_preparation}) the normalization is satisfying.

We used pass-bands sizes of $\times$1.5, $\times$3, $\times$4.5 the width of the stellar line, where the Voigtian fit gives a stellar width of 0.75 {\AA}. We also selected the 16 {\AA} spectral window used in \citet{Jensen2012_Halpha}. Measurements through such band-passes results in $-0.123\pm0.012\%$, $-0.029\pm0.010\%$, $-0.035\pm0.008\%$,  and $0.140\pm0.027\%$ in the $\times$1.5, $\times$3, $\times$4.5, and 16 {\AA} pass-bands, respectively.

Top of figure~\ref{fig:Halpha} shows only one significant signal detected in the narrower pass-band. However, this signal need carreful interpretation because even if the transmission spectrum (bottom fig.~\ref{fig:Halpha}) shows an unusual feature in the bluer wing and an absorption near the core of stellar H$\alpha$, this absorption match with the telluric spectrum and therefore wrongly corrected telluric variations at water vapour transitions location may produce a spurious signal.

For HD 209458b, our signal detected in the $\times1.5$ (1.125 {\AA}) band is equivalent to a radial extent of $3928\pm397$km (7.1$\times$H). For the $Ly\alpha$ transition \citet{Vidal-Madjar2003_lymanAlpha} obtained $15\pm4\%$ absorption using a central pass-band between 1215.15 {\AA} and 1216.1 {\AA}. In comparison, for HD 189733b  \citet{Jensen2012_Halpha} report an $H\alpha$ detection at a level of $-0.0872\pm0.0148\%$ using a 16 {\AA} pass-band.

\subsection{On the Ca I and Sc II detection feasibility}

Clausius-Clapeyron equation\footnote{$P_{Vap}=P_{Vap,0} \,\ exp \left( -\frac{\Delta H}{R} \,\ \left[ \frac{1}{T}-\frac{1}{T_0} \right] \,\ \right)$, where $P_{Vap}$ is the vapour pressure, $\Delta H$ is the latent heat of the phase transition, $R$ is the specific gas constant and $T$ the temperature.} may be useful to analyse the phase diagram of a chemical element through the temperature and pressure. Altitudes ($z_{Sc, Ca}$) derived previously for the absorbents in the atmosphere were used to obtain the atmospheric pressure-temperature conditions for HD 209458b \citep{Vidal-Madjar2011_sodium_TP, Vidal-Madjar2011_sodium_TP_Corrigendum}, where we considered isothermal layers\footnote{$P=P_0 \,\ exp \left( - \frac{z_{Sc, Ca}-z_{min}}{H} \right)$, where $P$ is the pressure and $H$ es the scale height}.

\begin{figure}[t]
\centering
\includegraphics[scale=0.4]{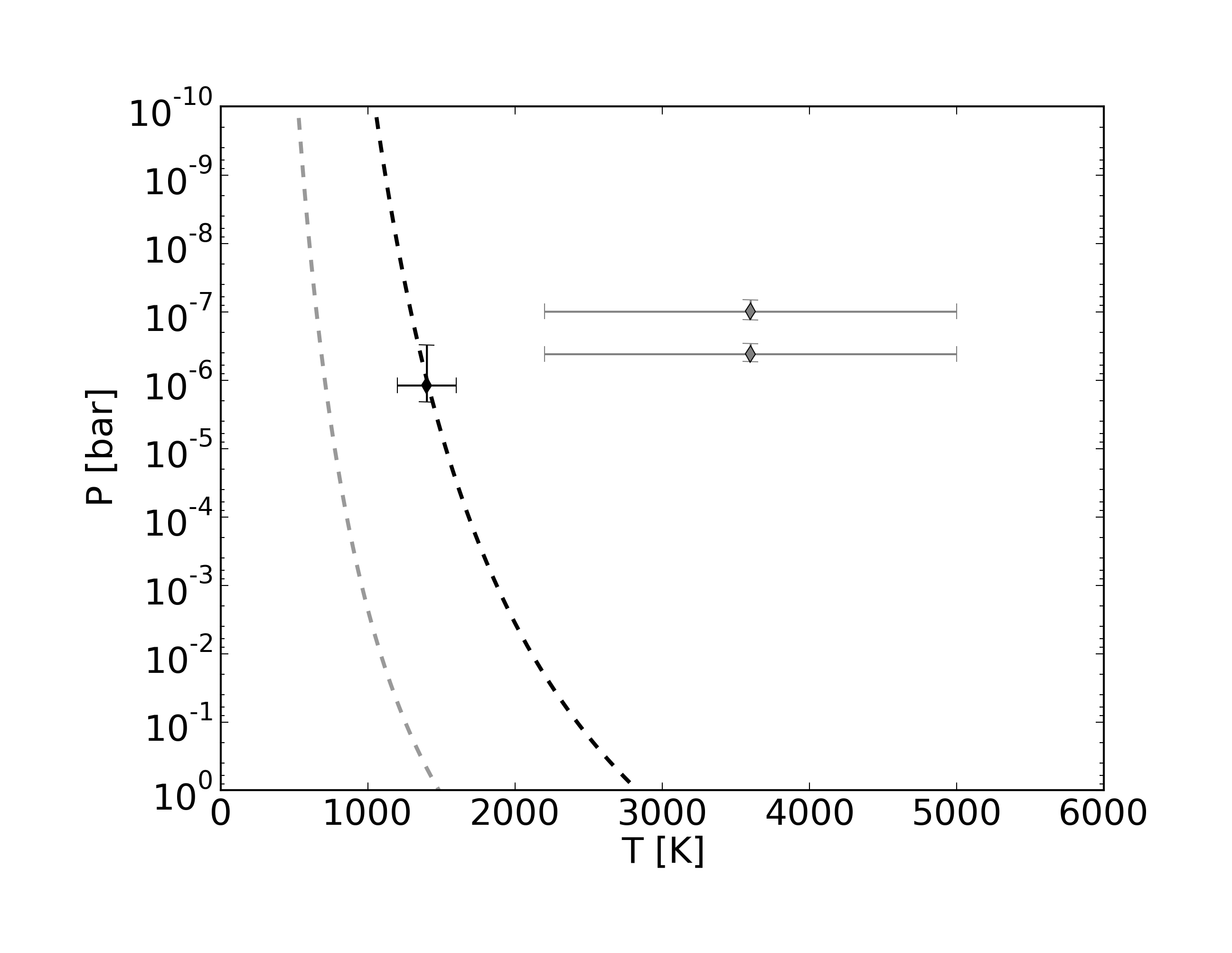}
\caption{Vapour pressure curves are represented by dashed curves. Red dashed curve represent the Calcium vapour pressure curve and the blue dashed curve represent the Scandium Vapour pressure curve. Diamonds shows the temperature and pressure conditions of Calcium (red) and Scandium (blue) derived from the measured altitudes.}
\label{fig:TP_VP}
\end{figure}

Calcium and Scandium physical parameters were obtained from \citet{2010crc..book.....L}. Figure~\ref{fig:TP_VP} shows the derived vapour pressure curves and pressure-temperature conditions derived from altitudes ($1896\pm379$ km for Sc II, and $2537\pm396$ km and $4419\pm429$ km for Ca I at 6162.27 {\AA} and 6493.78 {\AA}, respectively). We note that Calcium clearly remains in the gaseous phase while Scandium grazes the vapour pressure curve in the liquid state. Sc clouds could thus be present.

Atmospheres models have predicted condensation of Ca, but such condensation via $CaTiO_3$ does not seem to necessarily imply the vanishment Ca as TiO and O had been previously detected in the atmosphere of HD 209458b \citep{Vidal-Madjar2004_O_C_HD209458b, Desert2008_TiO_VO_HD209458b}.

\section{Conclusions}
We present the first ground-based detection of Calcium and possibly Scandium and Hydrogen in the transmission spectrum of HD 209458b. Ca I at 6162.17 {\AA} and 6493.78 {\AA}, Sc II transition at 5526.79 {\AA}, and $H\alpha$ presents an absorption excess of $(-7.9\pm1.2)\times10^{-4}$, $(-13.8\pm1.3)\times10^{-4}$, $(-5.9\pm1.2)\times10^{-4}$, and $(-12.3\pm1.2)\times10^{-4}$ in the $\times$3, $\times$3, $\times$4.5, and $\times$1.5 pass-band, respectively. Na D and Ca I show the expected rise of observed altitudes toward smaller pass-bands, while the Sc does not present such a behaviour. Although the $H\alpha$ signal is clear and shows stability through bootstrap analysis, the origin of such feature in the transmission spectrum is not completely clear.

HD 209458b is highly irradiated exoplanet where atmospheric conditions allow Ca to be in the gaseous state given observed conditions in \citet{Vidal-Madjar2011_sodium_TP, Vidal-Madjar2011_sodium_TP_Corrigendum} and our observed altitudes. According to Clausius-Clapeyron analysis Sc could condensate forming clouds or hazes.

We note that even if models expect the weakening and vanishing of Ca, TiO, VO, and O - e.g. via CaTiO$_3$, see section 3.3 in \citet{Jensen2012_Halpha}, \citet{Allard2012_Atmospheres_VLM_EP} - the atmosphere of HD 209458b present TiO, VO, and O \citep{Vidal-Madjar2004_O_C_HD209458b, Desert2008_TiO_VO_HD209458b} and here we present Ca detection.

We demonstrated the potential of automatic and ``blind'' search of exoplanetary atmospheres evidences in the transmission spectrum resulting in the detection of species that, at the present time, have been not predicted by theoretical models.

Additionally, \citet{Fossati2010_Wasp12B_metals} detected metals in the atmosphere of WASP-12b, Sc among them. These results put in concordance the fact that HD 209458b and WASP-12b are highly irradiated hot-Jupiters and they may present a similar photochemistry.

We also validate a method to perform telluric correction using science data itself, without the need of a reference star or models. We rely on the validity of such method is limited to data acquired in one stable single night as it involves a minimum of variation in the contents of Earth's atmospheric components.

\begin{acknowledgements}
N. A-D. acknowledges support from GEMINI-CONICYT FUND and from Programa Nacional de Becas de Posgrado (Grant D-22111791). P. R. and N. A-D. acknowledges partial support from FONDECYT \#11080271 and \#1120299. The authors gratefully acknowledge the valuable comments of the anonymous referee.
\end{acknowledgements}

\bibliographystyle{aa}
\bibliography{CaI_HD209458b_vFinal}

\end{document}